\documentclass{aa}

\usepackage{graphicx}
\usepackage{array}
\usepackage{tabularx}
\usepackage{amsmath}
\usepackage{multirow}
\usepackage{natbib}

\usepackage{hyperref}
\bibpunct{(}{)}{;}{a}{}{,} 

\usepackage{comment}
\usepackage[dvipsnames]{xcolor}
\usepackage[normalem]{ulem}
\usepackage[export]{adjustbox}
\hypersetup{colorlinks=true, citecolor=blue}
\usepackage[all]{hypcap}
\usepackage{orcidlink}

\newcommand{\msun}{M$_{\sun}$}
\newcommand{\rsun}{R$_{\sun}$}
\begin{document}
  \title{Sub-stellar engulfment by a main sequence star:\\ where is the lithium?}
  \titlerunning{Sub-stellar engulfment by an MS}

  \author{R. M. Cabez\'on\inst{1,2} \orcidlink{0000-0003-3546-3964}
  \and
  C. Abia\inst{3}
  \orcidlink{0000-0002-5665-2716}
  \and
  I. Dom\'\i nguez\inst{3}
  \orcidlink{0000-0002-3827-4731}
  \and
  D. Garc\'ia-Senz\inst{4,5}
  \orcidlink{0000-0001-5197-7100}
  }

\institute{Scientific Computing Center (sciCORE), Universit\"at Basel, Klingelbergstrasse 61, CH-4056 Basel, Swizerland\\\email{ruben.cabezon@unibas.ch}
\and
 Departement Physik, Universit\"at Basel, Klingelbergstrasse 82, CH-4056 Basel, Swizerland
\and
 Departamento de F\'\i sica Te\'orica y del Cosmos. Universidad de Granada, E-18071 Granada, Spain
\and
 Departament de F\'{i}sica, Universitat Polit\`ecnica de Catalunya, Av. Eduard Maristany 16, E-08019 Barcelona, Spain
\and
Institut d'Estudis Espacials de Catalunya, Gran Capit\`a 2-4, E-08034 Barcelona, Spain
}

  \date{Received ; accepted }

  \abstract
  {Since the discovery of exoplanetary systems, questions have been raised as to those sub-stellar companions that can survive encounters with their host star, and how this interaction may affect the internal structure and evolution of the hosting star. In particular its surface chemical composition.}
  {We study whether the engulfment of a brown dwarf (BD) by a solar-like main-sequence (MS) star can alter significantly the structure of the star and the Li content on its surface.}
  {We perform 3D Smoothed Particle Hydrodynamics simulations of the engulfment of a BD with masses 0.01 and 0.019 M$_\sun$, onto an MS star of 1 M$_\sun$ and solar composition, in three different scenarios: a head-on collision, a grazing collision with an impact parameter $\eta=0.5$~R$_\sun$, and a merger. We study the dynamics of the interaction in detail, and the relevance of the type of interaction and the mass of the BD on the final fate of the sub-stellar object and the host star in terms of mass loss of the system, angular momentum transfer, and changes in the Li abundance in the surface of the host star.}
  {In all the studied scenarios most of the BD mass is diluted in the denser region of the MS star. Only in the merger scenario a significant fraction ($\sim 40\%$) of the BD material would remain in the outer layers. We find a clear increase in the surface rotational velocity of the host star after the interaction, ranging between 25~km~s$^{-1}$ (grazing collision) to 50~km~s$^{-1}$ (merger). We also find a significant mass loss from the system (in the range $10^{-4}-10^{-3}$ M$_\sun$) due to the engulfment, which in the case of the merger, may form a circumstellar disk-like structure. Assuming that neither the depth of the convective envelope of the host star nor its mass content are modified during the interaction, a small change in the surface Li abundance in the head-on and grazing collisions is found. However, in the merger we find large Li enhancements, by factors 20-30, depending on the BD mass. Some of these features could be detected observationally in the host star provide they remain long enough time.}
  {In our 3D simulations, a sizeable fraction of the BD survives long enough to be mixed with the inner core of the MS star. This is at odds with previous suggestions based on 1D simulations. In some cases the final surface rotational velocity is very high, coupled with enough mass loss that may form a circumstellar disk. Merger scenarios tend to dilute considerably more BD material on the surface of the MS star, which could be detected as a Li-enhancement. The dynamic of the simulated scenarios suggests the development of asymmetries in the structure of the host star that can only be tackled with 3D codes, including the long-term evolution of the system.}


  \maketitle

\section{Introduction}
\label{sec:intro}
The enhanced depletion of lithium in the Sun, discovered almost 70 years ago \citep{gre51},
remains the epitome of the solar Li puzzle. The surface Li\footnote{Throughout this paper we will refer as lithium (Li) to the most abundant of its isotopes $^7$Li.  Photospheric abundances are indicated on the 12-point scale as A(X)$=\rm{log(N_{X}/N_{H}) + 12}$, where N$\rm{_{X}}$ and N$\rm{_{H}}$ represent the number of atoms of species X and H, respectively.} abundance in the Sun \citep[A(Li)$=1.05\pm 0.10$ dex,][]{aps09} is about 140 times less than the initial protosolar abundance, which is assumed to be the meteoritic value \citep[A(Li)$\sim 3.30$, e.g.][]{lod19}. Nevertheless, we note that very recently, on the basis of a new 3D non-thermodynamic equilibrium analysis, \citet{wan21} finds a Solar surface Li abundance of A(Li)$=0.96\pm 0.05$ dex.
However, the base of the surface convective layer of the Sun is not hot enough for nuclear reactions to destroy Li, which is destroyed by proton and alpha capture reactions at temperatures higher than $\sim 2.5\times 10^6$ K. This Li abundance anomaly is not unique to the Sun; large dispersion in Li abundances is observed in solar-type main sequence (MS) stars of similar age, mass, and metallicity \citep[see, e.g.][]{lam04}, which is difficult to explain within the standard models of stellar evolution.
Canonical stellar evolution predicts MS stars with Li-depleted interiors \pagebreak[4] \citep[see for example][]{dan94,pia02,dan03,mon06}, but almost no surface depletion during the MS, nor the dispersion observed in MS stars with very similar stellar parameters. As a consequence, some non-standard mixing mechanisms have been proposed to explain this observational result such as rotation, mass loss, diffusion, and/or gravity waves \citep[e.g.][]{pin92,swe92,mic86,ume00,char05}, but a satisfactory solution has not been found yet. This has opened the possibility of an external origin as the cause for this depletion. Tidal interaction with sub-stellar objects like brown dwarfs (BD) and/or planets, or even their engulfment are some of the most frequent alternative hypotheses. This latter scenario would not be uncommon, as planetary engulfment seems to be the fate awaiting a significant number of planetary systems. In fact, transit surveys indicate that hot Jupiters (HJ)\footnote{Hot Jupiters are defined as gas giant planets with a period of $\sim$ 10~days and an orbital separation $\leq 0.1$~AU.} are found orbiting $\sim$ 1\% of sun-like stars \citep[e.g.][]{win15}.

\citet{ale67} first suggested that the Li depletion observed in exoplanet-host stars could be directly related to the mechanism of formation and/or migration of planets in the corresponding stellar system. This idea has been explored in many observational studies \citep[see e.g.][]{kin97,gon00,che06,isr09,gon10,del15,gon15}.
The potential discrepancy between lithium abundances in MS stars hosting or not exo-planets is, however, controversial and still divides the observer community. Most of the arguments against the existence of this discrepancy rely on a wrong interpretation of observational data or on an observational bias \citep[e.g.][]{rya00,luc00,bau10}. The reason for this is, in part, the difficulty in finding a large enough homogeneous sample of MS stars having equal age, mass, metallicity, rotational velocity, etc. Furthermore, the derived Li abundances in many of these observational studies are not accurate enough to extract any solid conclusion.

A way to partially overcome this issue is carrying out detailed differential chemical analyses in solar-like twin stars in wide binary systems under the hypothesis that they were formed at the same time, in the same birth cloud with nearly identical composition.
Nowadays, this type of analysis is able to determine chemical abundances with unprecedented (relative) precision ($\lesssim 0.01$ dex for some elements, see e.g.
\citet{oh18,liu18,mai19,nag20}).
Unfortunately, the planet population around most of these solar-like twin stars is unknown, and in the few ones where planets have been detected, these may orbit or not (apparently) the star showing a chemical anomaly. As far as we know, in only three of the solar-like binary systems studied so far, Li has been measured accurately enough: the pairs 16 Cyg A-B, Kronos/Krios, and HIP 71726/HIP 71737 \citep{yan21,mai19,oh18}. Surprisingly, Li is found to be more abundant in the most metal-rich star of the pair (for instance in 16 Cyg A with respect to B), the star which presumably accreted rocky material (i.e. a sub-stellar object). This seems to contradict the conclusion from the Li surveys in planet-host stars mentioned above, which suggests an extra Li depletion in the host star. Therefore, from the observational point of view, the impact of the engulfment of a sub-stellar object by an MS star on the surface Li abundance remains still unclear.

From the theoretical side, planetary accretion may occur from the pre-main sequence through the white dwarf stage \citep[see e.g.][]{kai21,kru21}, and different physical processes drive the accretion at different stages \citep{jac18}. Nevertheless, only a few studies exist in the literature on the interaction between an MS star and a sub-stellar object \citep[see e.g.][]{met12,yam17,jia18}, since the overwhelming majority of them are devoted to the interaction occurring during the cool giant phases of the stellar evolution, in which the probability of planetary engulfment increases \citep{sie99a,sie99b,vil14,pri16,pri16b,agu16,sta16,jac18,mac18,agu20,soa20}.
Most of these theoretical studies are one-dimensional (1D) analytical simulations and usually assume that: a) the accreted sub-stellar object or rocky-material dissolves completely in the stellar envelope at the point where the Virial temperature of the accreted material equals that in the envelope \citep{sie99a}, and b) the engulfment does not alter significantly the internal structure and evolution of the host star. Obviously, the former hypothesis is critical for evaluating the impact on the surface chemical composition of the hosting star. On this basis, \citet{mon02} first studied the planet accretion onto solar-type stars and showed that it may significantly change the surface abundance of Li if it takes place during the main sequence evolutionary stage. Nevertheless, they concluded that there is a narrow range of stellar masses and metallicities where Li can be efficiently preserved in the envelope. This was contradicted in part by \citet{bar10} who found that episodic accretion onto young solar-type stars can produce objects with significantly higher central temperatures than non-accreting stars of the same mass and age. As a consequence, Li depletion is larger in these objects. This could explain the extra Li depletion observed in some young stars belonging to open clusters. Later on, \citet{the12} and \citet{ddd15} showed that the accretion of rocky material onto the host star creates an inverse gradient of molecular weight, which leads to thermohaline convection. As a consequence, heavy elements are mixed downwards until the mean molecular weight gradient becomes nearly flat. In most cases, a tiny or no signature of the accreted heavy elements remains at the surface after a few Ma. They concluded that the accretion of heavy matter cannot lead to any increase of the Li abundance at the surface, but conversely, it may lead to a reduction.

\begin{table*}
\caption{Summary of the simulations setup, stating the name of the model, the mass, number of SPH particles (N), the initial separation ($d_0$), and the impact parameter ($\eta$). The last column shows the approximate total CPUh used for each calculation. Models A correspond to $M_{BD}=0.019$~\msun, while models B correspond to $M_{BD}=0.01$~\msun. Models 00 are head-on collisions (zero impact parameter), models 05 are grazing collisions with impact parameter $\eta=0.5$~\rsun, and models M are mergers.}
\centering
\begin{tabular}{@{}|c|cc|cc|c|c||c|@{}}
\hline
\multirow{2}{*}{Model} & \multicolumn{2}{c|}{MS star} & \multicolumn{2}{c|}{BD star} & $d_0$ & $\eta$ & \multirow{2}{*}{CPUh}\\
                       & Mass (\msun) & N & Mass (\msun) & N & (\rsun) & (\rsun) & \\
\hline
\hline
A00 &   &   &  &  & 2.0 & 0.0 & 155,000\\ 
A05 & 1.00 & 5,260,000 & 0.019 & 100,000 & 2.0 & 0.5 & 168,000\\ 
AM &    &   &  &  & 1.5 & - & 585,000\\ 
\hline
B00 &    &   & &  & 2.0 & 0.0 & 204,000\\
B05 & 1.00 & 5,260,000 & 0.01 & 52,600 & 2.0 & 0.5 & 175,000\\
BM &    &   &  &  & 1.5 & - & 524,000\\
\hline
\end{tabular}
\label{table_models}
\end{table*}

To our knowledge, the sole 3D studies on the accretion of a sub-stellar object by an MS star are those by \citet{san98,san02}. These authors performed hydrodynamic simulations to study the giant-planet (Jupiter/Saturn-like planets) consumption by a solar-like host star and, as a consequence, its possible Li pollution. They showed that depending on details of the planetary interior, partial or total dissolution of giant planets can result in significant enhancements in the metallicity of the host star with mass in the range of $1.0\lesssim$M/\msun$\lesssim 1.3$. Nevertheless, their calculations had a very limited spatial resolution and did not simulate the full structure of the MS star. Additionally, their simulations lacked a mechanism to transport angular momentum from the planet to the star's envelope.

From the discussion above it follows that both, observational and theoretical studies, are far from being conclusive and many questions still remain open in the engulfment scenario. What is the fate of the sub-stellar object? Does it survive, or is it destroyed by ablation or tidal disruption? If so, in which fraction? Will the interaction lead to an alteration of the MS star properties, such as spin-up, mass-loss, luminosity, size, or a change in the surface abundances, in particular, that of Li?

In the present study, we perform 3D hydrodynamical simulations of the engulfment of a BD into an MS star of $\sim 1$~\msun~and solar composition by using the smoothed particle hydrodynamics (SPH) method. Our main aim is to answer some of the questions quoted above mainly concerning the final fate of the sub-stellar object and the consequences that the engulfment may have on the host star in terms of its mass loss, rotation, and in particular on the possible change of the surface Li abundance. We focus our study on the relevance of two parameters: the mass of the BD and the type of interaction. In a forthcoming study, we will address the same issues but changing the evolutionary status of the host star to a red giant branch and/or horizontal branch star. The structure of the paper is the following: in Section~\ref{sec:numerical} we describe the numerical setup and the simulated engulfment scenarios together with the main approximations that were made. Section~\ref{sec:results} discusses the results of these simulations, particularly the impact that the engulfment provokes on the global properties of the hosting star and on its surface Li abundance. Finally, in Section~\ref{sec:conclusions} we discuss the main results, present our conclusions, and hint at directions to continue this study.

\section{Numerical setting}
\label{sec:numerical}
Traditionally, SPH codes have been used to simulate planetary impacts and stellar collisions \citep{slattery92, Allison97} due to their intrinsic conservation properties and ability to follow complex geometries. To simulate the scenarios presented in this study, we used the hydrodynamic code SPHYNX \citep{cabezon2017}. SPHYNX\footnote{\url{https://astro.physik.unibas.ch/sphynx}} is a state-of-the-art SPH code, that includes most of the latest upgrades in the SPH technique, including the Integral Approach to Derivatives \citep{garcia2012,cabezon2012,rosswog2015,valdarnini2016}, the $sinc$ family of interpolating kernels \citep{cabezon2008}, artificial viscosity switches \citep{cullen2010,read2012}, and generalized volume elements \citep{saitoh2013,hopkins2013,gar2022}. It also includes a two-level dynamic load-balancing implementation that considerably reduces the load imbalance that might appear in a distributed calculation \citep{mohammed2020}. Overall, this code is well-suited to cope with the highly dynamic, geometrically distorted evolution of stellar collisions, and the development of hydrodynamical instabilities, even in the subsonic regime. The ability of the SPH technique to handle free surfaces is also particularly suitable for this scenario, where we can simulate both full stars and follow the fluid that may escape the system. Additionally, the spatial resolution is adaptive (which in SPH translates to variable smoothing length), increasing where density increases and reaching resolutions as high as $10^8$~cm (i.e. $\sim 1.4\times10^{-3}$~\rsun) in our configuration. Finally, SPH codes have excellent energy and momentum conservation properties, which is of critical importance when evaluating angular momentum transfer in dynamic scenarios such as those studied here.
\begin{figure*}
\includegraphics[width=\textwidth]{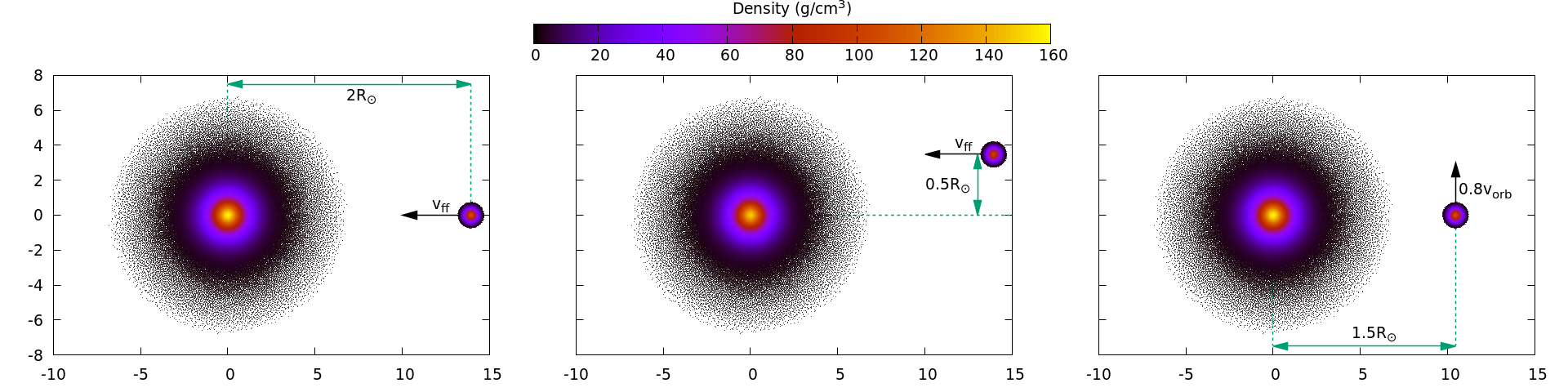}
    \caption{Onset of each simulated scenario: head-on collision (model A00, left panel), grazing collision (A05, middle), and merger (AM, right). We show here the particles in a thin slice along the orbital plane (X-Y plane) for models A. Models B have the same setup, correspondingly. Axes are in units of $10^{10}$~cm and density is color-coded.}
    \label{fig:initialpos}
\end{figure*}
\begin{figure*}
\includegraphics[width=\textwidth]{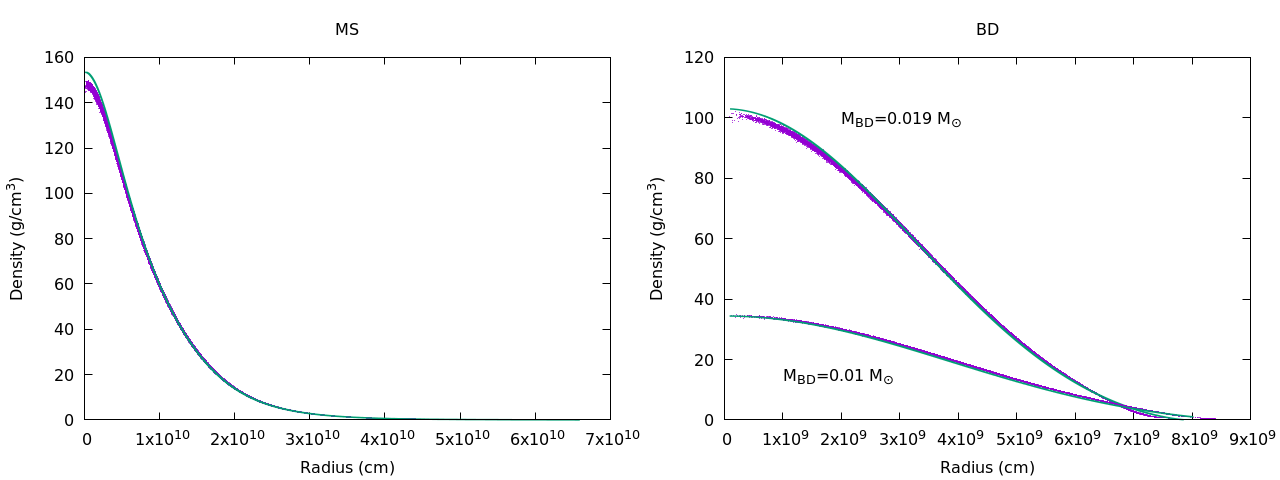}
    \caption{Density profiles for the MS (left panel) and BD (right panel) stars after relaxation. All SPH particles are represented with purple dots. Green solid lines show the initial 1D theoretical density profiles.}
    \label{fig:initialmodels}
\end{figure*}

\subsection{Simulated scenarios}
\label{subsec:scenarios}

In Table~\ref{table_models} we list the scenarios that we simulate. These are divided into two groups depending on the mass of the BD: models A correspond to $M_{BD}=0.019$~\msun, while models B use $M_{BD}=0.01$~\msun. With the former mass choice, we want to maximize the possible effects of the interaction of a sub-stellar object onto the host star and, with this relatively low BD mass we ensure that the original Li content in the BD is fully preserved. We note that previous 1D studies - although using simplifying assumptions - revealed the existence of a critical mass below which the colliding sub-stellar object is evaporated within the envelope. This mass limit was found to be $\sim 0.02$~\msun~\citep[e.g.][]{liv84}, hence we are in the position to test this result in a 3D setup. The second choice is close to the theoretical mass limit threshold separating a planet from a BD, $\sim 13$~M$_J$ \citep[see e.g.][]{spi11}. On the other hand, the idea behind using the lower-mass BD model is to get closer to the planet engulfment scenario but avoiding the need to simulate rocky structures, which may imply changing the equation of state.
For both models (A and B), we explore three types of interactions: a head-on collision, a grazing collision, and a merger. The former two are more violent, have shorter timescales, and are less realistic than the latter due to their extremely low cross section. Therefore, we expect to see significant differences in the outcome of these interactions. Models 00 and 05 in Table~\ref{table_models} are collisions: head-on and grazing, respectively. Models M, which probably represent a more realistic scenario, are mergers induced via setting the BD in an orbit around the MS star with an orbital velocity artificially reduced by $20\%$ to induce the merger\footnote{Strictly speaking, a merger is also a collision, but we will reserve the term \emph{collision} for the head-on and grazing interactions.}. The last column of Table~\ref{table_models} shows the approximate amount of CPUh employed for each calculation. These values are purely indicative. Despite being all models calculated with 224 cores, the simulations were performed in different facilities with different generations of CPUs and network interconnections. The wall-clock time for each simulation was 30-40 days for the collisions and 100 days for the mergers.

Figure~\ref{fig:initialpos} shows a snapshot of the initial positions of the stars for each different interaction. Note in Table~\ref{table_models} the factor $\sim 52$ difference in the number of particles (i.e. SPH fluid elements) between the BD and the MS in models A, and about a factor of 100 for models B. This is due to the fact that we use the same mass for all the particles to avoid unwanted numerical effects arising from the interaction between particles with different masses. 

\begin{figure*}
\includegraphics[width=\textwidth]{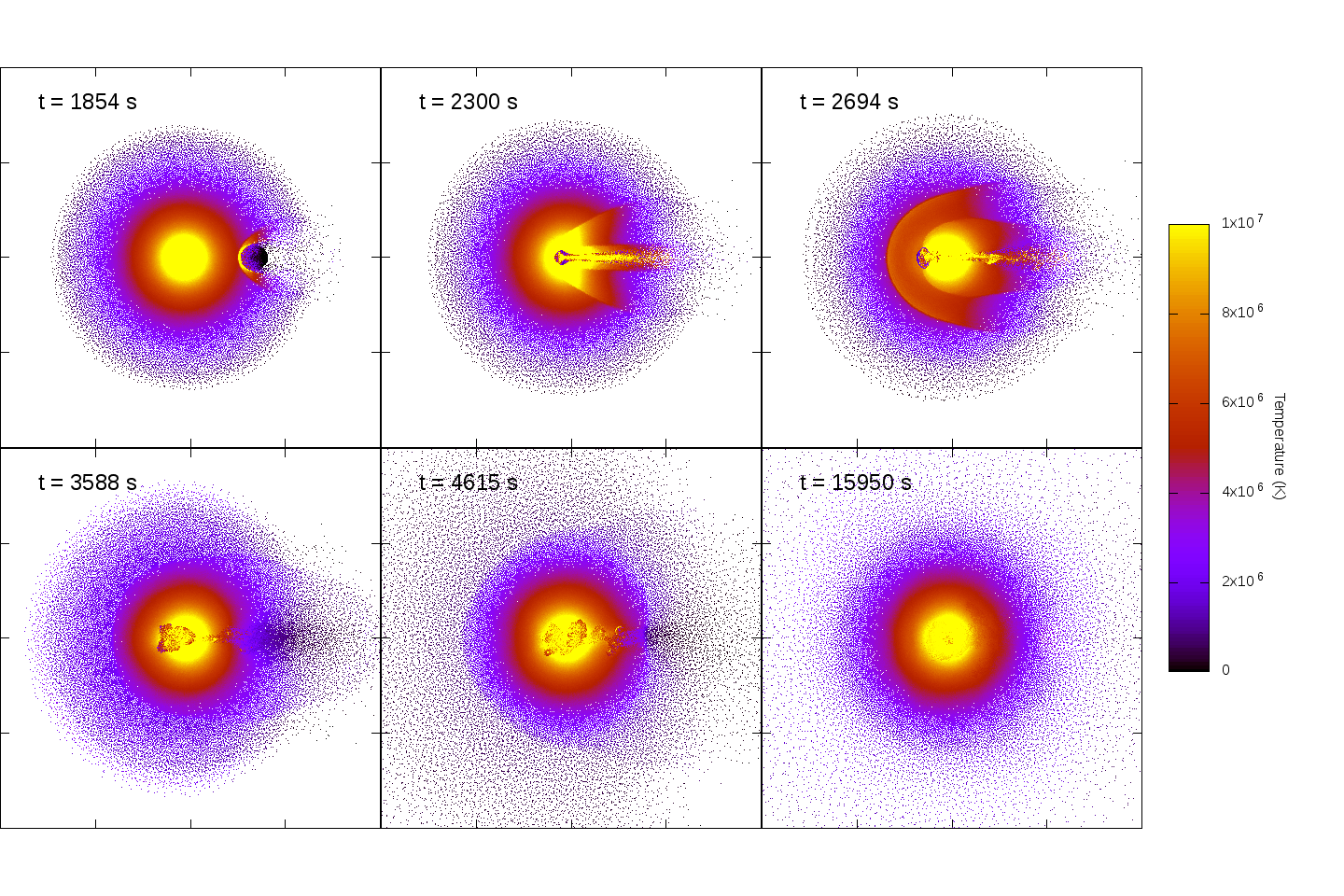}
    \caption{Snapshots of the particle distribution for the head-on collision scenario with $\eta=0$ (model A00) at different times, increasing from left to right and from top to bottom. We only show particles within a thin slice along the orbital plane. Each snapshot shows a box with side $2\times10^{11}$~cm and temperature is color-coded.}
    \label{fig:snapshots_a00}
\end{figure*}

\subsection{Initial conditions}
\label{subsec:ICs}
The initial configuration of the MS star, which mimics the current Sun, is achieved by distributing randomly SPH particles in 3D, following 1D density, temperature, and composition profiles of the Sun at age $\sim 4.5$ Ga from \citet{bah05}.
According to this solar model the bottom of the convective envelope is located at $r\sim 0.71$ R$_\odot$ and contains a mass of $\sim 0.024$ M$_\odot$.
After relaxing the star, we obtain a configuration that is very close to the initial 1D profiles. We used a stellar Helmholtz equation of state (EOS), which includes the contributions of radiation in the form of Planckian photons, an ideal gas of ions with Coulomb corrections, and an electron-positron gas with an arbitrary degree of relativistic regimes and degeneracy \citep{timmes2000}. This EOS has a range of applicability wide enough to allow us to simulate both the MS, the BD, and their interaction. It is important to stress that while the MS is close to an ideal gas, the BD is mostly supported by non-relativistic degenerate electrons. Having the same EOS for both interacting bodies renders unnecessary any complex treatment of interacting/mixing matter coming from different stars. This would not be the case if the secondary object were a planet.

Finally, we adopted a lithium surface abundance (i.e. in the convective envelope) similar to that observed on the surface of the current Sun (A(Li)$=0.96$), while in the interior of the MS star the Li abundance profile is that for the current age of the Sun obtained in a full evolutionary calculation of a 1 M$_\odot$ and solar composition stellar model. The computed profile shows that the Li abundance becomes negligible immediately below the convective envelope. We note that the depth and mass of the convective envelope do not change significantly from the zero-age main sequence to the current Sun. The relative evolution of the depth of the convective envelope and the total solar radius is less than $0.2\%$ per $10^9$ a, while the convective envelope mass would decrease from the initial 0.0329 M$_\odot$ to the current value of $0.02415$~\msun~\citep{bah01}. Therefore, if the engulfment occurred during the very early main-sequence evolution, the outcome would not be very different.

To obtain the initial model of the BD we integrate the mechanical equilibrium configuration using the Helmholtz EOS. At each integration step, we calculate the temperature profile assuming that the degeneracy parameter ($\eta$) is constant, fixed with the central density ($\rho_c$), temperature ($T_c$), and composition (via the mean molecular weight per electron, $\mu_e$),

\begin{equation}
    \eta_c=\frac{(3\pi^2\hbar^3)^{2/3}}{2m_ek_BT_c}\left(\frac{N_a\rho_c}{\mu_e}\right)^{2/3} \sim \frac{3.018\times 10^5}{T_c}\left( \frac{\rho_c}{\mu_e}\right)^{2/3}
\end{equation}

\noindent where the $c$ subscripts mean central values and all other constants have their usual meaning. Hence, the temperature radial profile is obtained as:
\begin{equation}
    T(r) = \frac{\eta(r)}{\eta_c}T_c
\end{equation}
This approximation is valid for most of the BD structure for a given mass, composition, and elapsed time, as the degeneracy parameter is approximately constant along adiabats \citep{burrows1993}. This method produces reasonable density profiles for brown dwarfs given the use of a detailed EOS, such as the Helmholtz EOS used in this work. Although the resulting temperature profile is only qualitatively correct, especially close to the surface of the BD, the details of the initial temperature profile are not too relevant owing to the strong heating caused by the collision.

\begin{figure*}
\includegraphics[width=\textwidth]{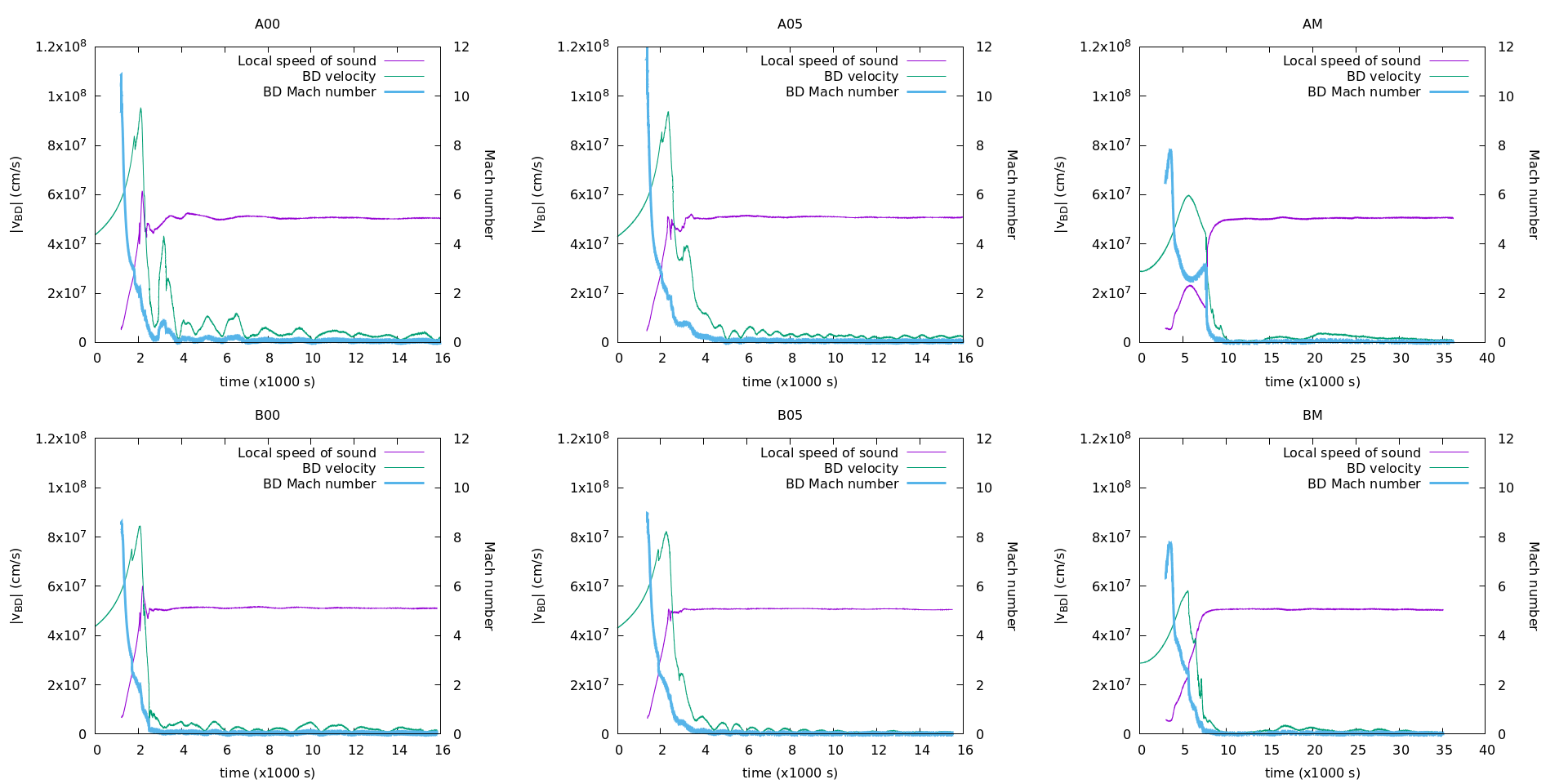}
    \caption{BD velocity evolution (green line) compared with the local speed of sound of the surrounding MS star's material (purple line). The BD Mach number evolution is shown in blue. The top panels show models A, while the bottom panels are for models B.}
    \label{fig:velevol}
\end{figure*}

In a second step, we map these profiles into a 3D random distribution of SPH particles and relax the system letting it achieve a stable configuration. Being the BD a long-lived, cold, and fully convective star, we consider a constant Li abundance throughout it and equal to the accepted current value in the interstellar medium (ISM), A(Li)$\sim 3.0$. We used $100,000$ (models A) or $52,400$ particles (models B) for the BD so that all SPH particles in the simulation have the same mass. Figure~\ref{fig:initialmodels} shows the initial density profile for the corresponding objects. All SPH particles are shown as dots jointly with the corresponding 1D theoretical density profile as solid lines. The low dispersion of the SPH profiles and their similarity to the 1D profiles show that the 3D initial conditions adequately represent the theoretical profiles.

Once we have the initial configurations of the individual objects, we can arrange the onset of the interaction. Many studies addressed the physics of star-planet interactions, some of them focusing on how a planetary orbit changes under the action of tides between the star and the planet \citep[see e.g.,][]{liv84,sie99a,sie99b,vil07,nor10,bea11,vil14,pri16,pri16b,ste20}. These studies have provided many interesting clues about the initial conditions for an engulfment to occur in terms of the mass of the star, mass of the planet, initial distance between the planet and the star, rotation, and of the importance of other physical ingredients such as the mass loss rates or convective overshooting. However, the details of the previous orbiting evolution of the system and the ultimate reason that may have caused the collision is beyond the scope of this study. Therefore, depending on the simulated scenario, we set both objects apart, with an initial distance $d_0$. Then, we assign an initial velocity $v_{0,rel}$ to both, which corresponds to the free-fall velocity ($v_{ff}$) at $d_0$ for the collisions, and $80\%$ of the orbital velocity ($v_{orb}$) for the merger. Figure~\ref{fig:initialpos} shows the configurations with respect to the center of the MS for simplicity. Nevertheless, the simulations are performed with respect to the center of mass frame of reference, in which both bodies have non-zero initial velocities.
\begin{figure*}
\includegraphics[width=\textwidth]{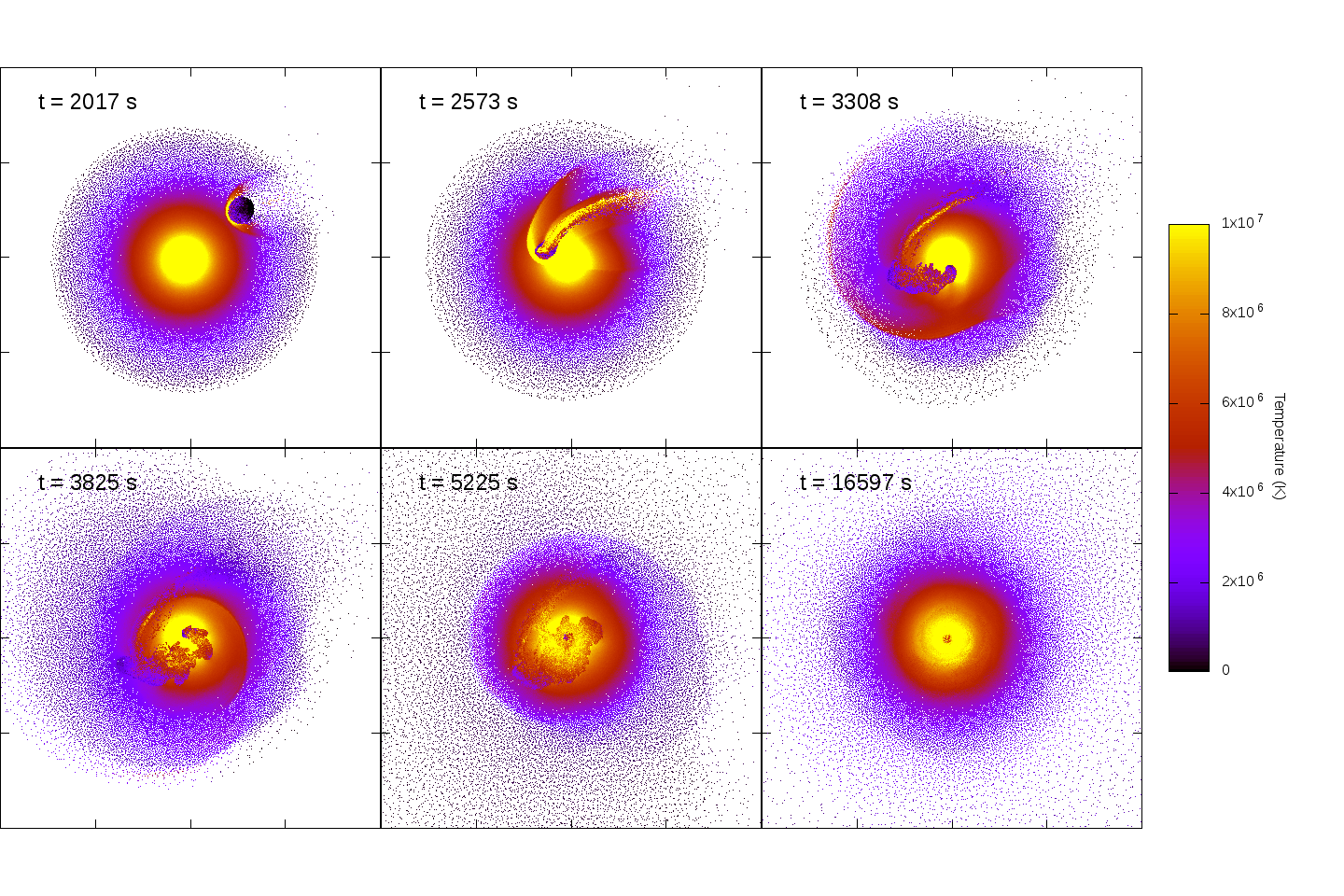}
    \caption{Snapshots of the particle distribution for the grazing collision scenario with impact parameter $\eta=0.5$ (model A05) at different times, increasing from left to right and from top to bottom. We only show particles within a thin slice along the orbital plane. Each snapshot shows a box with side $2\times10^{11}$~cm and color represents temperature.}
    \label{fig:snapshots_a05}
\end{figure*}
\section{Results}
\label{sec:results}
In this section, we present the results obtained in our simulations. Using cases A (i.e. $M_{BD}=0.019$~\msun) as nominal scenarios, we first focus on the dynamics of the system, exploring in detail the (more extreme) cases of head-on and grazing collisions, and finally the smooth merger. Secondly, we discuss the mixing and distribution of BD material after the engulfment, particularly how its Li content is distributed in the host star\footnote{We refer to the stellar object indistinctly as MS or host star, hereafter.}. Thirdly, we comment on the angular momentum transfer in the case of a grazing collision and the merger scenario. Finally, we discuss the effects of reducing the mass of the BD (cases B, $M_{BD}=0.01$~\msun).

\subsection{Analysis of the dynamics}
\label{subsec:dynamics}

\subsubsection{Head-on collision}
\label{subsec:eta00_dynamics}
Model A00 represents our nominal case, corresponding to a head-on collision with zero impact parameter. In the following, we describe the evolution of the collision as it resulted from our 3D simulation.


Figure~\ref{fig:snapshots_a00} shows several snapshots taken at different times during the collision (model A00). Figure~\ref{fig:velevol} (left) shows the velocity evolution of the denser region of the BD in comparison with the local speed of sound of the surrounding MS star material. Shortly, after first contact, a bow-shock is formed as the BD penetrates the MS star supersonically (with a peak Mach number of $\mathcal{M}\simeq11$) and it speeds up towards the core of the MS star. After a short ballistic phase, the BD losses part of its mass by a combination of direct mass-stripping and ablation. Nevertheless, it manages to cross the core of the MS star, while it heats up and compresses. Due to the extreme friction experienced by the BD, its kinetic velocity is being leveraged to heat the material. This slows down the BD, which is not able to break out through the MS star antipodes, making the BD move subsonically until its direction is inverted ($t\simeq2700$~s). At this point, the fragmentation of the BD increases the ratio surface-to-volume, which makes it prone to dissolve. Mixing of the BD material with that of the MS star is facilitated by the Kelvin-Helmholtz instability, which has a characteristic growth time of
\begin{equation}
    \tau_{KH}=\frac{\lambda(\rho_1+\rho_2)}{\sqrt{\rho_1\rho_2}\left|\mathbf{v}_{rel}\right|}\,.
\end{equation}

\noindent
with $\rho_1$ and $\rho_2$ being the densities of the interacting fluids, $\lambda$ is the size of the growing mode of the KH instability, and $\mathbf{v}_{rel}$ is the relative velocity. Assuming that $\rho_1\sim\rho_2$, $\mathbf{v}_{rel}\sim5\times10^7$~cm~s$^{-1}$, and $\lambda\sim R_{BD}$ being the mode responsible of the BD destruction, we can estimate $\tau_{KH}\sim350$~s. According to numerical simulations of the destruction of a homogeneous sphere by a supersonic wind, approximately $t\sim3\tau_{KH}$ is needed to fully destroy the sphere \citep[see e.g.][]{agertz2007,hopkins2013,frontiere2017,gar2022}. This pushes our estimation for the destruction of the bubble at the order of $t\sim1000$~s after the bowshock is formed ($t\sim1800$~s in Fig.~\ref{fig:snapshots_a00}). Namely, $t\sim2800$~s. In our simulation, neither the density of the BD nor the MS star is constant, nor is the relative velocity of the fluid while the BD traverses the MS star, which results in longer survival of the BD $t\sim3500$~s, but still similar to our estimation. At this point, we cannot talk anymore of a BD \emph{per se}, but of blobs of fluid that previously belonged to the BD. Thanks to the Lagrangian nature of SPH, we can follow the fluid elements of the MS star and BD independently, even in the case of extreme mixing. In that respect, the remaining fragments of BD are not in dynamical equilibrium, and they experience again a gravitational pull towards the center of the MS star. This time they completely dissolve by the combined effect of hydrodynamic instabilities and heat losses through their surface, effectively mixing with the MS star. All these mixing episodes take place in relatively high-density regions of the MS star, keeping the fresh material of the BD in deeper layers and bound to sink to the center of the star. We could see two repetitions of this process (see the first two peaks in the BD velocity curve of Fig.~\ref{fig:velevol}) until the BD material is completely mixed with the MS star.
A shockwave is launched during the first crossing episode of the BD, which eventually breaks through the outer layers of the MS star, inducing some amount of mass ejection (see Table~\ref{table_results}). To calculate a lower limit to the mass lost by the system we added the mass of all SPH particles whose radial velocity was positive and greater or equal to the escape velocity evaluated at their radius, considering the total enclosed mass and correcting by the position and velocity of the center of mass of the MS. About $1.9\times10^{-3}$~\msun~are ejected from the system. Of those, only $7.8\times10^{-6}$~\msun~were originally from the BD. This amount of ejected mass is quite similar to the total mass in the Solar System, excluding the Sun.

\begin{table*}
\caption{This table summarizes the outcomes of our simulations. The columns show the model name, the BD mass that has $r>0.7$~\rsun~and is still gravitationally bounded at the end of our simulations, the amount of total mass lost by the system, the amount of mass lost by the system that corresponds to the BD, the period of the fluid with $r<$\rsun~at the end of the simulation, respectively.}
\centering
\begin{tabular}{@{}|c|cccc|@{}}
\hline
\multirow{2}{*}{Model} & BD mass [$r>0.7$~\rsun]& Ejected total mass & Ejected BD mass & Period \\
                       & ($\times 10^{-4}$~\msun) & ($\times 10^{-4}$~\msun) &  ($\times 10^{-6}$~\msun) &  (days) \\
\hline
\hline
A00 & 2.58 & 19.54 & 7.79 & 351.11 \\
A05 & 0.68 & 22.35 & 0.95 & 1.11 \\
AM & 78.93 & 4.85 & 0.00 & 0.97 \\
\hline
B00 & 2.39 & 5.93 & 23.38 & 199.03 \\
B05 & 0.93 & 6.38 & 0.19 & 2.04 \\
BM & 39.90 & 6.30 & 0.00 & 1.74 \\
\hline
\end{tabular}
\label{table_results}
\end{table*}

\subsubsection{Grazing collision}
\label{subsec:eta05_dynamcis}
We also performed a variant of the previous scenario, now with a non-zero impact parameter (model A05 with $\eta=0.5$~\rsun). This scenario is more likely than a head-on collision of the BD, and it may modify the angular momentum of the MS star which, could be detected observationally (see Sec.~\ref{sec:angmom} below).

Figure~\ref{fig:snapshots_a05} shows snapshots of the distribution of the particles at different times, with the temperature being color-coded. The initial part of the interaction is similar to the A00 case. The BD impacts supersonically (peak $\mathcal{M}\simeq 12$) against the outer layers of the MS star, forming a bow-shock that crosses the entire host star. The BD is compressed as it gets closer to the denser core of the host star. Nevertheless, due to the non-zero impact parameter, the BD misses the center of the MS star and spirals around it. In this case, it takes a longer time than case A00 for the BD to be able to slow down. As a consequence, a sizeable fraction of the initial spiral trajectory of the BD happens supersonically (see the top central panel in Fig.~\ref{fig:velevol}). This launches a spiral shockwave during the first part of the interaction. When the BD reaches more diluted regions of the MS star,  it expands and triggers the growth of hydrodynamical instabilities during the next part of the orbit, that mix part of its material (up to $70$\%) with that of the MS star. The remaining material of the BD reaches the inner core of the MS star half of an orbit later and slows down. At the end of the simulation, about $27$\% of the BD material is in the inner core of the MS, most of the BD material ($45$\%) is set in a dense toroidal structure around the MS core and a remaining $28$\% is in a tight spiral arm. From here on, both structures are slowly pulled in by the gravitational force of the MS core. As in the A00 case, most of the interaction happens within relatively deep layers of the MS star, hence most of the mass of the BD will end up in the densest regions of the host star.

The outer layers of the host star expand as they receive the impact of the spiral shockwave and a slightly higher amount of material is ejected from the system than in the case of A00 ($\sim 2.2 \times 10^{-3}$~\msun). In this case, the amount of BD mass that is lost is almost one order of magnitude smaller than in the A00 model ($\sim 9.5\times 10^{-7}$~\msun) for reasons that are explained in Sect.~\ref{subsec:Li_abun}. It is worth noting the low-temperature material that can be seen at the center of the MS core in the last panel of Fig.~\ref{fig:snapshots_a05}. This is just a transitory situation. Once the BD loses kinetic energy, its material will thermalize with that of the MS core, but on a longer timescale than that simulated in this work. Therefore, despite its low temperature, this material will be, on all counts, burned anyway.

\subsubsection{Merger}
\label{subsec:merger_dynamics}
\begin{figure*}
\includegraphics[width=\textwidth]{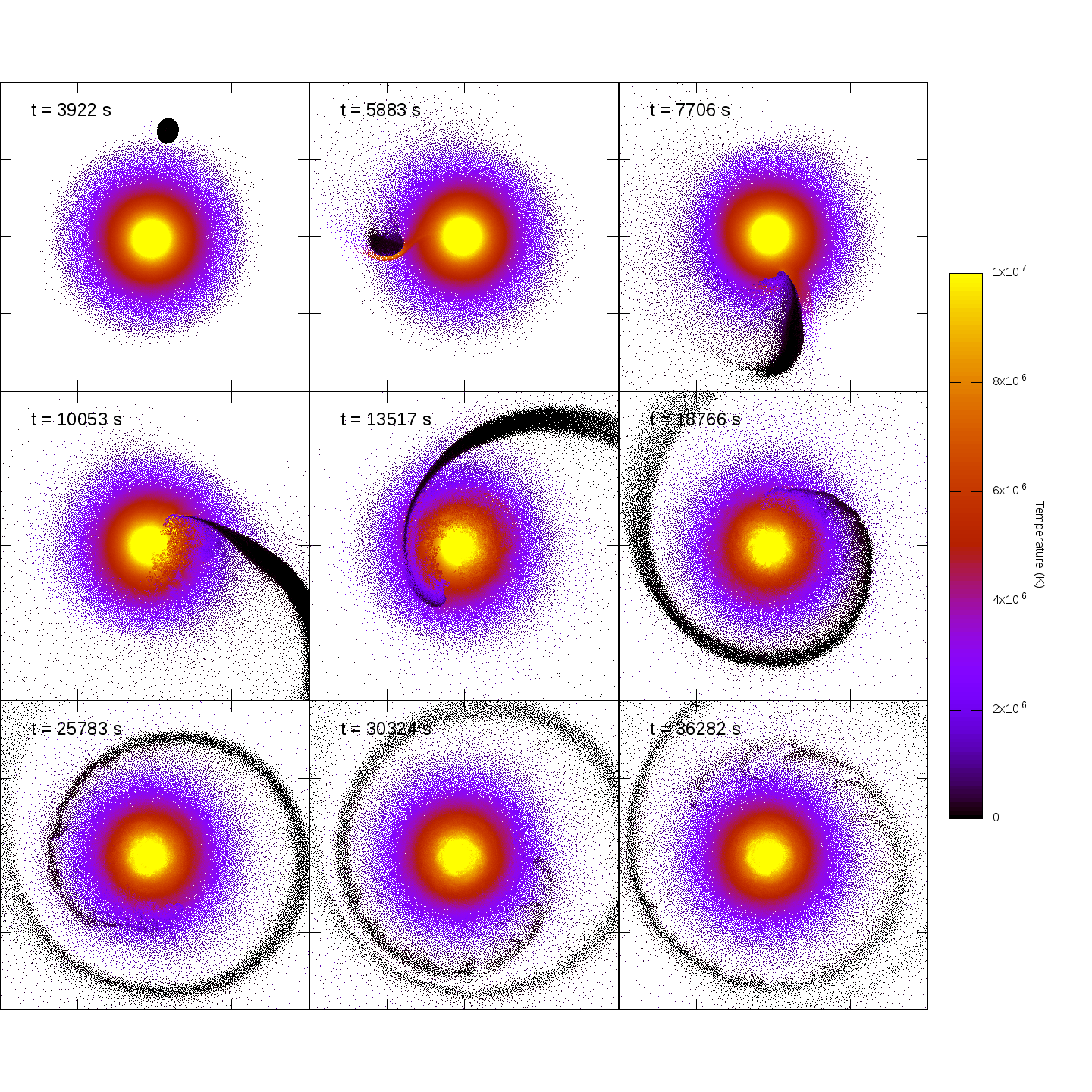}
    \caption{Snapshots of the particle distribution for the merger scenario (model AM) at different times. We only show particles within a thin slice along the orbital plane. Each snapshot shows a box with side $2\times10^{11}$~cm and color represents temperature.}
    \label{fig:snapshots_am}
\end{figure*}
As mentioned in Sec.~\ref{subsec:scenarios}, in order to induce the merger of the BD, we set it in an orbit ($d_{orb}=1.5$~\rsun) with an orbital velocity artificially reduced ($20\%$ less than the velocity needed for a stable orbit at that distance). This leads the BD into a trajectory that gets it close to the external layers of the MS star in a quarter of an orbit. Figure~\ref{fig:snapshots_am} shows the evolution of the merger at different stages with color-coded temperatures. The first interaction with the material of the MS star happens supersonically ($\mathcal{M}\simeq 8$) and due to the relatively low density in the outer layers of the MS star, the BD is effectively sped up by the gravitational potential of the host star during the next half orbit. Part of the BD material ($\sim 10$\%) is stripped out via ablation and it mixes with the outer layers of the MS, but the stronger effect is due to tidal forces. By the time when the BD reached the antipodes of the entry point, it has been noticeably distorted and stretched (third panel of the top row in Fig.~\ref{fig:snapshots_am}). Due to the combined work of the tidal forces and the brief slingshot acceleration, the BD is stripped into a long spiral arm that is partially accreted on top of the MS core while the remaining material orbits around the MS (central row of Fig.~\ref{fig:snapshots_am}).

As the interaction proceeds, the spiral tightens and keeps accreting on the outer layers of the MS (bottom row of Fig.~\ref{fig:snapshots_am}) until about 60\% of the BD mass is in the innermost regions of the MS and the remaining 40\% in the spiral arm. The final state of this simulation is still not stationary (last snapshot in Fig.~\ref{fig:snapshots_am}), but the timescale needed in order to allow the spiral arm to settle down on the MS surface is much longer than what can be affordable in a simulation limited by the dynamical timescale.

Nevertheless, the amount of the total mass lost by the system is smaller than in the more violent head-on and grazing collisions: $\sim 5 \times 10^{-4}$ M$_\sun$ (see Table~\ref{table_results}), and no material belonging to the BD escapes the system, meaning that the full spiral arm will eventually fall back. Figure~\ref{fig:BD_dist} (right panel) shows that this material would form a circumstellar disk in the plane of the collision with possible observational consequences in terms of infrared excess, provided that this structure lasts for a long enough time.

\subsection{Li abundance}
\label{subsec:Li_abun}
In our scenarios, the BD fluid elements act as tracers of Li-rich material. Because of the Lagrangian nature of SPH, we can directly track where the BD material is located throughout the whole simulation, simply by following the SPH particles of the BD. Determining where this material ends within the MS star structure allows us to predict the expected observable Li abundance after the collision or merger. Admittedly, we can only simulate several thousands of seconds with a full 3D hydrodynamics code, as the collision is very dynamic and we are limited by the Courant condition. To give a final answer on where the Li-rich material ends, it would be necessary to continue this calculation up to millions of years with a 3D stellar evolution code, which nowadays is impossible. Due to the highly axial asymmetry of the final outcome of the simulations, the common procedure of mapping the final 3D structure into a shell-averaged 1D profile and following the evolution with a spherically symmetric hydrostatic code is not a trivial task. Nevertheless, we can provide some constraints on the fate of the BD material and guidelines on what to expect from scenarios such as those simulated here.

\begin{figure*}
\includegraphics[width=\textwidth]{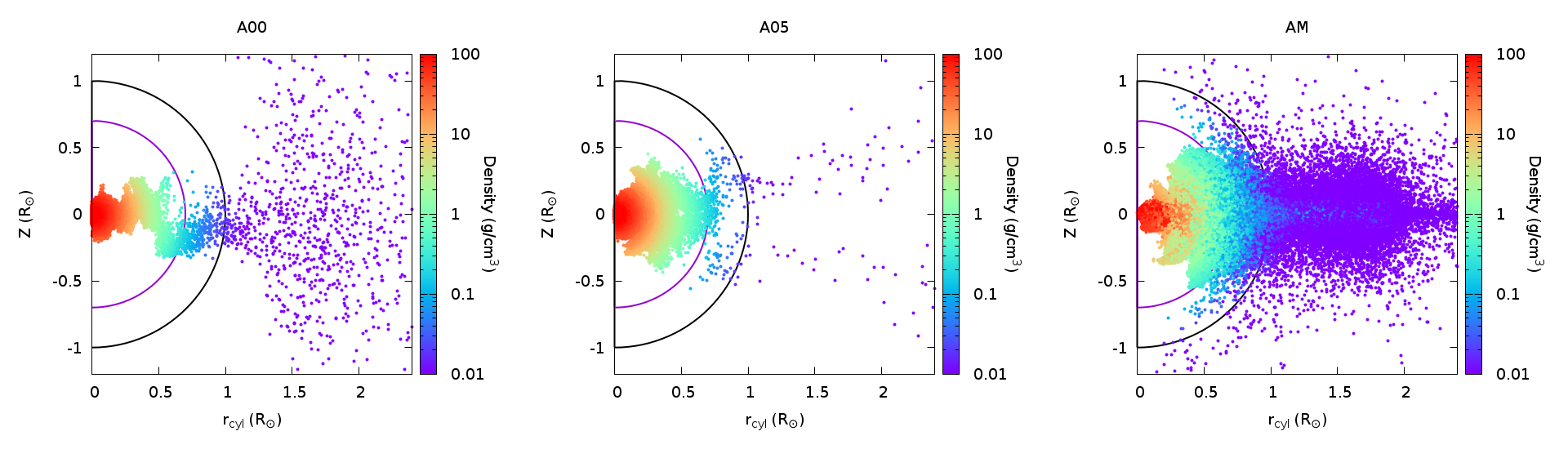}
    \caption{Particle distribution of the BD projected in the 2D cylindrical plane for the scenarios A, at the end of the simulations: $t=10689$~s and $t=8059$~s in the head-on and grazing collision cases (left and central, respectively) and $t=38402$~s for the merger scenario (right). The semicircles represent the original solar radius (solid black line) and the bottom of the convective zone at $\sim 70\%$ of the solar radius (purple solid line). Density is color-coded.}
    \label{fig:BD_dist}
\end{figure*}
\begin{figure*}
\includegraphics[width=\textwidth]{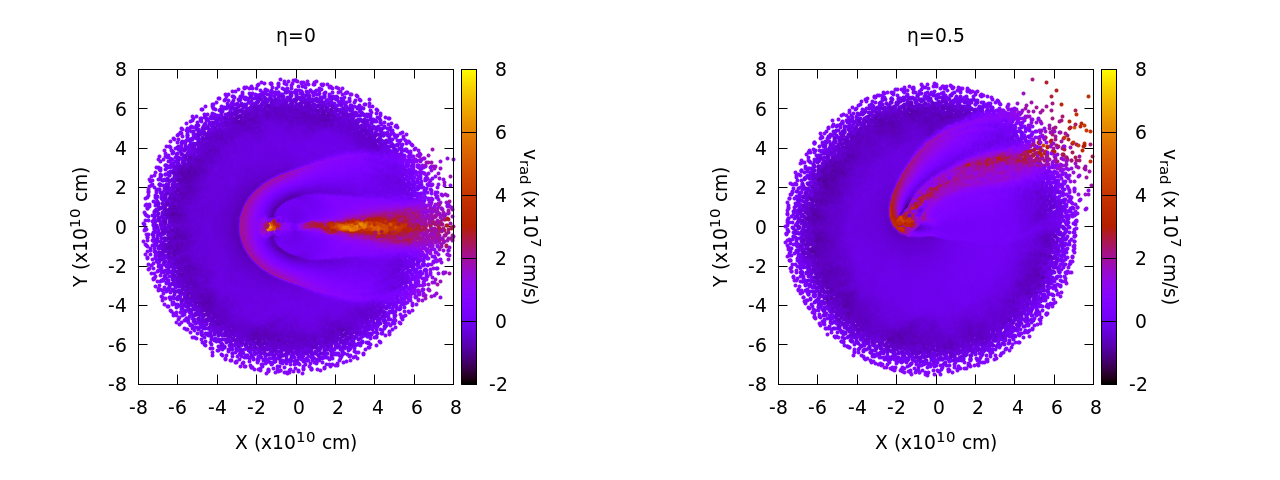}
    \caption{2D cut of the 3D particle distribution for cases A00 (left) and A05 (right) at $t=2600$~s. Radial velocity is color-coded.}
    \label{fig:funnel}
\end{figure*}
\begin{figure}
\includegraphics[width=\columnwidth, center]{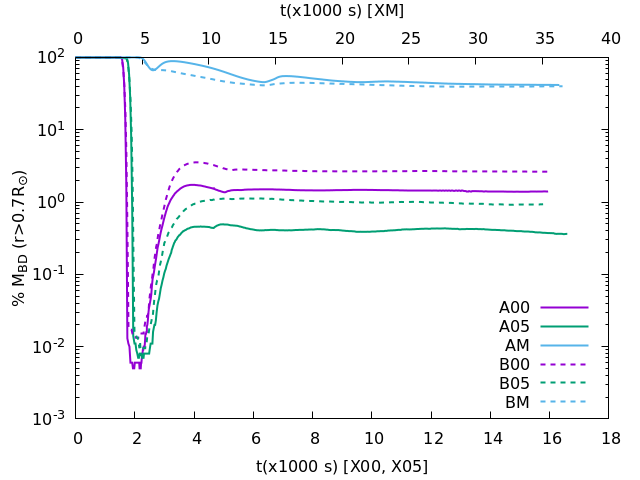}
    \caption{Time evolution of the percentage of gravitationally-bound BD mass with $r\geq0.7$~\rsun. Lower X-axis is for collisions, while the upper X-axis is for mergers. Solid lines are for models A and dashed lines are models B.}
    \label{fig:BDmass_07rsun}
\end{figure}
\begin{figure*}
\includegraphics[width=\textwidth]{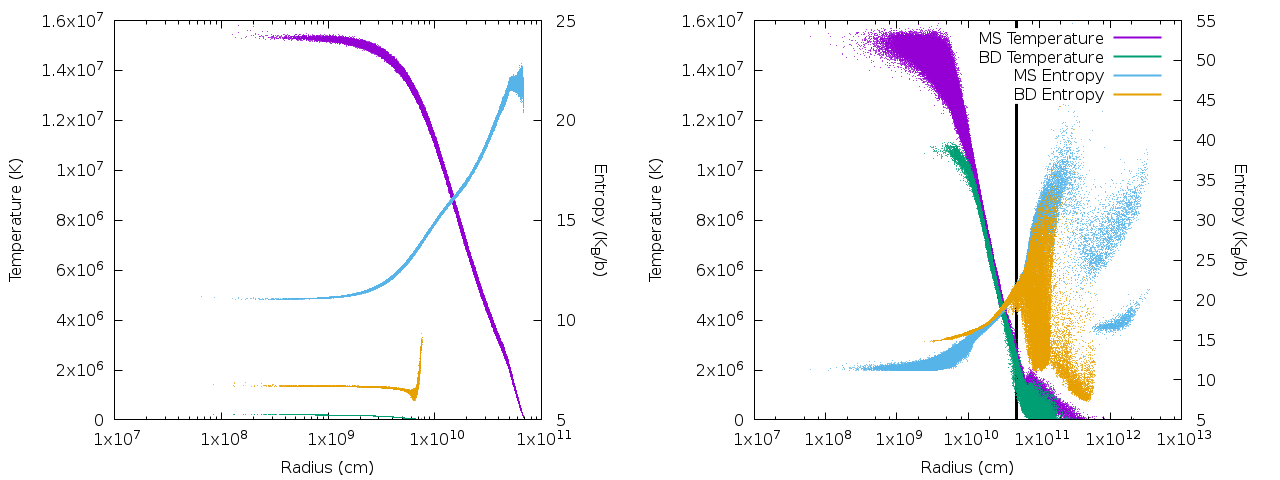}
    \caption{Initial (left) and final (right) temperature and entropy radial profiles for the MS and BD in the AM case. We show all SPH particles in all curves. The profiles in the left panel are centered in the center of mass of each star, while in the right panel they are centered in the center of mass of the remnant. The black vertical line in the right panel show the bottom of the convective envelope.}
    \label{fig:entropy}
\end{figure*}
We start discussing how the MS and BD materials are spatially distributed at the end of the simulations. Figure~\ref{fig:BD_dist} shows a cylindrical projection of all SPH particles of the BD at the end of each simulation of cases A. The first noticeable detail is that the BD material is not homogeneously distributed after the collision. There is a higher concentration along the equatorial plane (namely, the collision plane) than in the polar plane. This density inhomogeneity may have an impact on the subsequent evolution of the MS star. However, this issue must be addressed with a 3D stellar evolutionary code, which is beyond the scope of the present study. As mentioned above, the final outcome of our simulations has a fully inhomogeneous non-axisymmetric structure in density, temperature, and composition. Therefore, the extraction of any averaged 1D structure has the risk of losing the information regarding the collision into the MS star. For the cases of head-on and grazing collisions, most of the BD material is confined in the deep interior of the MS star, while only a small fraction remains above the original convective zone and is still gravitationally bounded ($1.32\%$ and $0.35\%$~for the A00 and A05 models, respectively), which we assume remains unperturbed at $r\sim 0.7$~\rsun, as well as the original stellar mass above this point ($\sim 0.02$ M$_\odot$, see also Table~\ref{table_results}) after the collision has settled down to its final configuration. As we cannot follow the evolution of the final object until it reaches equilibrium\footnote{Thermal equilibrium is expected to be achieved in a much longer time, of the order of the Kelvin-Helmholtz characteristic time for the MS star, $\tau_{KH}\simeq 10^7$~a.}, the location of the new convective envelope might be different. This rough factor two of difference in surface pollution between both scenarios can be understood when looking in detail at the dynamics of each collision. The A00 case has a clear preferred direction of movement for the ablated material of the BD: the impact direction. As the BD carves supersonically its path through the MS star's material, it leaves behind a temporary low-pressure funnel that allows the BD material to escape towards the MS star's surface. In the case of A05, because the BD spirals in, the low-pressure region behind the BD follows the same in-spiral path, away from the impact line, diluting any preferred direction for the BD matter to escape. This effect can be seen in Fig.~\ref{fig:funnel}, where we present a 2D cut along the collision plane for both collision scenarios of case A. The color of the particles represents their radial velocity. It is clear that a preferred direction with higher outward radial velocities exists in the case of A00, while the material moving outwards in the case of A05 has to push against layers of inert MS star's material.

\begin{figure}
\includegraphics[width=\columnwidth, center]{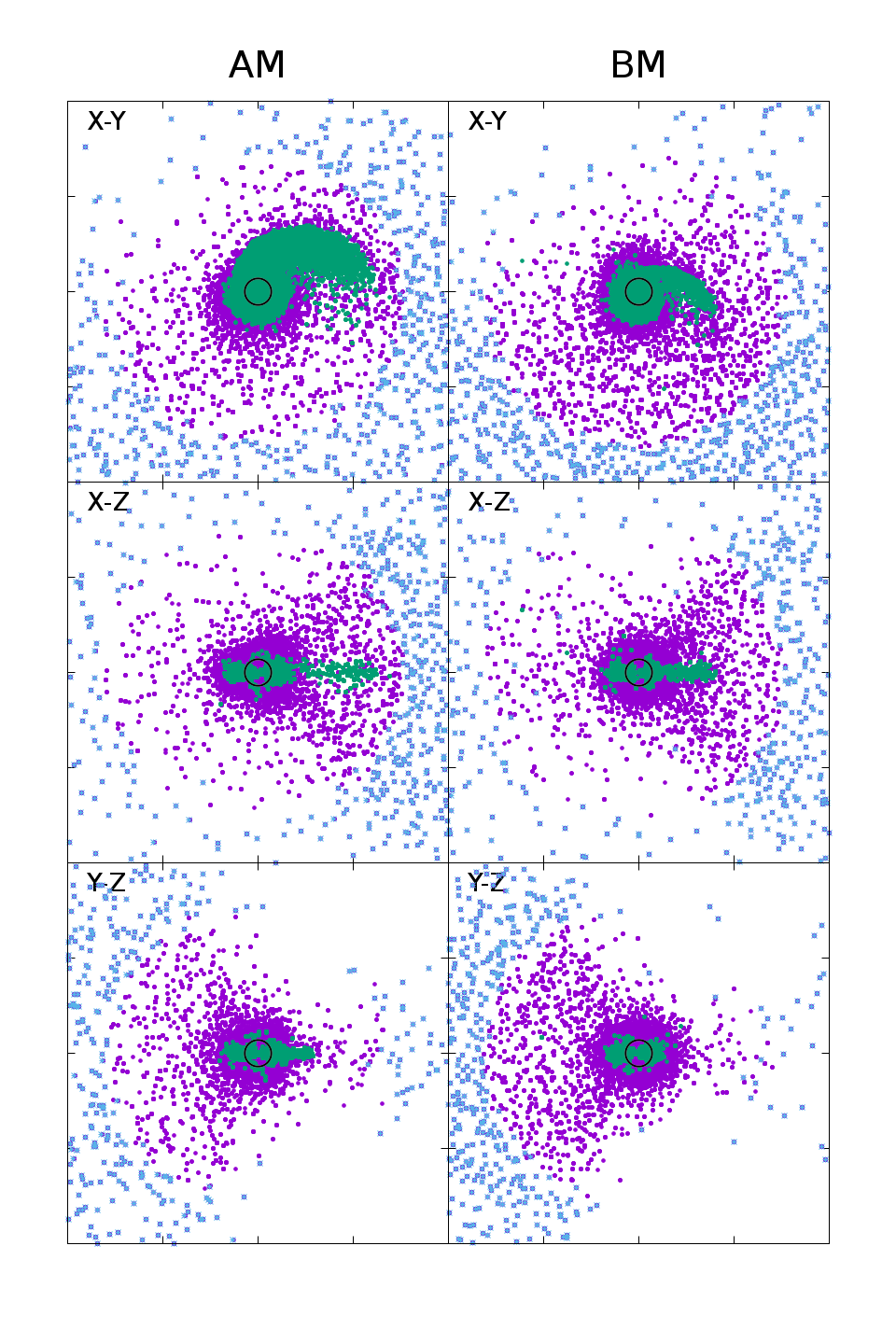}
    \caption{Spatial distribution for the MS star (purple) and the BD (green) material at the end of the merger scenarios AM (left column) and BM (right column). Particles in blue have radial velocity larger than the local escape velocity and all of them belong to the MS. Every row shows a thin slice in a different plane of the same 3D distribution. All panels have a size of $2\times 10^{12}$~cm per side and the black circle in the center of each panel shows the original radius of the MS star.}
    \label{fig:planarcuts}
\end{figure}

The case of the merger is completely different. As the BD spends a relevant amount of time in the diluted regions of the MS star, the ablated BD material remains in these outer layers and it never reaches denser and hotter regions. It is only when the BD moves closer to the MS core, that tidal forces are strong enough to rip the BD apart. Once this happens, the BD material is pulled in through a thin stream of matter, which due to angular momentum conservation, spirals around the MS core. The flow of BD material is not able to reach the central core of the MS, as it does in the collision cases. Nevertheless, it is able to deposit Li-rich material on the surface of the inner core of the MS, where it will be destroyed, nonetheless. A sizeable fraction ($\sim 40\%$) of the BD material is left on the remainder of the spiral arm, whose evolution happens on a longer timescale and it could eventually be allocated on the outer layers of the MS, within the convective region and its surface.

Figure~\ref{fig:BDmass_07rsun} shows the time evolution of the percentage of the BD mass that has radius $r\geq0.7~$\rsun~and is still gravitationally bound, for all simulated models. Solid lines correspond to models A. Here we can see that we reached a steady state on all simulations regarding the amount of BD mass that could eventually remain above the original MS convective region. The trend follows the results discussed above: head-on and grazing collisions have a lower capability to pollute the external layers of the host star than a merger.

It is important to note that in the merger case, about $50\%$ of the BD mass that has $r>0.7$~\rsun~ and is still gravitationally bound, shows an inversion of the entropy gradient. This inversion is the result of a mixture of ablated material that has been strongly heated by the shockwave during the first phases of the interaction and material expelled by simple angular momentum transfer during the destruction of the BD by tidal forces. Figure~\ref{fig:entropy} characterizes the significant heating that the BD material suffers, owing to the relatively high density of the environment, and its influence in its entropy profile. The panel on the left shows the initial profiles of temperature and entropy (in $k_B$ per baryon) for both stars. The panel on the right shows the final profiles for the same magnitudes. Note the change in temperature of the BD material (about two orders of magnitude closer to the MS star center) and the overall entropy increase in the outer layers of the MS star ($\sim\times2$) and the whole BD material ($\sim\times2-4$). In the latter, the entropy inversion is clearly seen for $r\gtrsim0.7$~\rsun, denoted by the solid vertical line. As a result, this entropy profile points to a convective unstable layer. The long-term effect of this instability is unclear, as the end state of all simulations is still too dynamic to predict the actual final location and state of those BD particles. Nevertheless, some mixing is expected to happen in a time-scale longer than what can be simulated with a 3D code nowadays. 

Figure~\ref{fig:planarcuts} shows the final coarse particle distribution of our merger simulations at a large scale. Matter from the MS is represented by purple points, while BD matter is in green. Blue points are the particles that will escape the system. As no matter from the BD reaches escape velocity (see ejected BD mass in Table~\ref{table_results}), it is clear from this plot that the final distribution of Li-rich material would be mostly located around the collision plane in an inhomogeneous way, forming a relatively thin spiral arm. Additionally, the outer layers of the MS have expanded considerably, although most of this matter is still gravitationally bound and will fall back. This figure gives a qualitative idea of how the Li-rich material is distributed at the end of the simulation, but not of its quantity, which was discussed in Sec.~\ref{subsec:merger_dynamics}.

\begin{figure*}
\includegraphics[width=\textwidth]{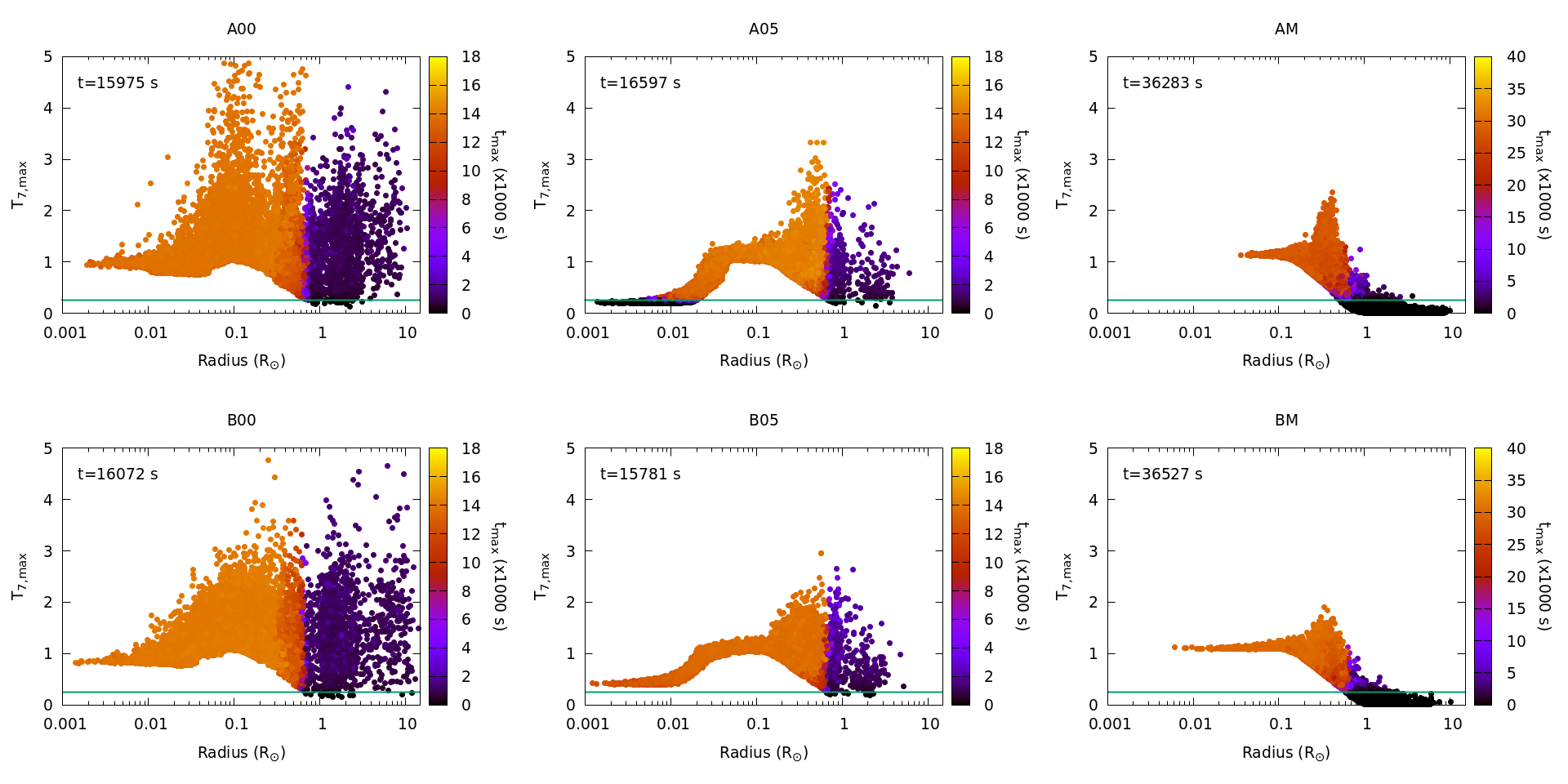}
    \caption{Maximum overall achieved temperature of the BD material in function of the radial distribution at the end of all simulated models. Color represents the amount of time that a specific particle had temperatures higher than $ 2.5\times 10^6$~K (i.e. the assumed Li burning temperature limit, denoted with a solid green line).}
    \label{fig:ttmaxTemp}
\end{figure*}

\begin{figure}
\includegraphics[width=\columnwidth, center]{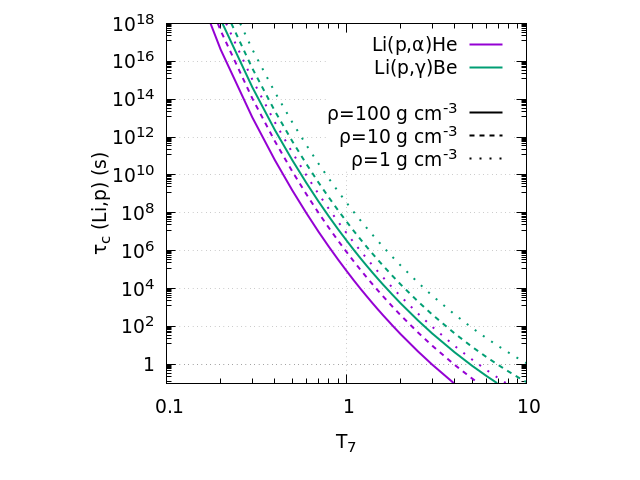}
    \caption{Characteristic timescales for the two main channels of Lithium burning ($^7$Li(p,$\alpha)^4$He in purple and $^7$Li(p,$\gamma)^7$Be in green) for different densities and in function of the temperature in 10$^7$ K units (T$_7$).}
    \label{fig:Liburntime}
\end{figure}

Besides the spatial distribution of the BD material, it is important to track their thermal evolution during the simulated time to assess if this matter ever reaches Li-burning temperatures ($\gtrsim 2.5\times10^6$~K) and if it does, for how long. To this extent, we tracked the maximum temperature of each BD SPH particle and the amount of time that they spend at temperatures larger than or equal to the previous temperature threshold. We show the results in Fig.~\ref{fig:ttmaxTemp}. The points show the maximum temperature ($T_{7,max}$, in 10$^7$~K units) reached by each particle of the BD throughout the whole simulation in function of its radial coordinate at the end of the calculation. Color represents the amount of time that a specific particle spent with temperatures higher than or equal to the Li burning temperature, denoted by the green solid line. In Fig.~\ref{fig:Liburntime} we show the characteristic burning time of Li for the two main channels of proton captures in function of temperature. The color of the lines represents the reaction channel and the type of line is for different density values that cover most of the range in our simulations.

In the collision cases, almost the full mass of the BD reaches temperatures much higher than $T_7=0.25$, but a fraction of those ($1.4\%$~of the BD mass for the A00 and B00 cases, and $0.5\%$ for the A05 and B05 cases) spends less than 1 hour with such temperatures and cools down below the Li-burning temperature limit within $1000-3000$~s. In particular, these include all those particles that end up at $r\geq 0.7$~\rsun. Cases 05 (middle panels of Fig.~\ref{fig:ttmaxTemp}) take a longer time to interact with the denser layers of the MS star and, consequently, the BD material reaches lower temperatures than the cases 00, and they spend less time above the Li-burning limit. This points to the fact that despite reaching very high temperatures, a fraction of the BD material is able to cool down relatively fast, and remains unburnt in the outer layers of the MS star. As mentioned in Sec.~\ref{subsec:eta05_dynamcis}, the decrease in temperature for the BD material at low radius is transitory, and this material is considered to be burnt in a longer timescale. Cases A05 and B05 show very similar behavior.


On the other hand, in the merger scenarios, almost $40\%$ of the BD mass never reaches the Li-burning temperature limit and an additional $4\%$ spends less than 1 hour at such temperatures. As it can be seen in Fig.~\ref{fig:ttmaxTemp} (right panel), most of this material remains at radius $\geq 0.7$~\rsun. The reason for this relies on the longer time that the BD spends interacting with the outer layers of the MS star. This interaction is enough to peel off material of the BD and develop a long thin spiral arm, which neither reaches the deeper layers of the MS star nor is substantially heated. The BD material in the merger simulations reaches minimum radii that are approximately one order of magnitude larger than in the collision scenarios. This suggests that merger scenarios are more prone to result in larger lithium contamination of the outer layers of the MS star.

On the other hand, all BD material with $r\leq 0.7$~\rsun~experiences temperatures much higher than $T_7=0.25$ for timescales comparable to the total simulated time in all simulations. This points to the fact that their lithium will be most likely completely burned in a relatively short timescale, as expected for the deeper regions of the host star. This limit between burnt and unburnt material, that can be clearly seen in Fig.~\ref{fig:ttmaxTemp} around $r\sim 0.7$~\rsun, independently of the simulated scenario, appears naturally. This led us to think that despite the grand display of material being ejected from the system and the BD being destroyed, the mass change is very low in comparison with the total mass of the MS star. Therefore, it is very likely that the location of the bottom of the convective envelope does not significantly change, hence our assumption of considering it still at $r\sim 0.7$~\rsun~throughout all the simulations.

Considering the above-quoted percents of the BD material which remains within the former convective region, it is straightforward to estimate the new Li surface abundance in the MS star after the collision/merger scenarios \citep[see e.g.][ their Eq.~10]{jac18}. As mentioned above, a sizeable fraction of this material is marginally exposed to temperatures higher than $T_7=0.25$ during the simulations (Fig.~\ref{fig:ttmaxTemp}). For the typical density and temperature values existing at the bottom of the convective envelope in the current Sun ($\sim 0.75$ g cm$^{-3}$ and $T_7\sim 0.2 $, respectively), the characteristic time for Li burning via the $^7$Li(p,$\alpha)^4$He reaction is $\sim 2\times 10^{11}$ a. Then, assuming the lithium abundance currently observed in the Sun as the surface abundance before the collision, the new surface abundance values would be A(Li)$=1.33$ and 1.09 for the A00 and A05 models, respectively. These increments in the absolute abundance of Li are small and it would be difficult to differentiate from the typical Li abundance observed in solar-like stars. However, in the case of the merger the Li enhancement is considerably larger due to the significant amount of the BD mass deposited in the envelope of the MS star: in the model AM the new Li abundance would increase by a factor larger than twenty  (A(Li)$\sim 2.46$). This huge increase in the Li abundance would be easily detectable observationally provided that this surface Li enhancement would remain for a long enough time. Note, that the trigger of thermohaline convection (see Sect.~\ref{sec:intro}) that eventually would erase this Li enhancement, would not develop in the scenarios studied here as the surface metallicity is not altered due to the engulfment; the MS star and the BD share identical metallicity only differing in the Li content.

\subsection{Angular momentum transfer}
\label{sec:angmom}
Except for the case of a head-on collision, all interactions should transfer angular momentum to the MS star. This is an important observable, which might hint at a history of engulfment. For instance, 1D simulations performed by \citet{car09,pri16,pri16b} show that the interaction between a red giant and a planet followed by planet engulfment can indeed accelerate the surface rotation for a sufficiently long time to produce fast rotating red giants with rotational surface velocities above $\sim 8$~km~s$^{-1}$. Rapid rotator red giants exist and usually show some Li-enhancement as compared to the slow rotator red giants, although the origin of this Li-enhancement is still unclear \citep[see e.g.][]{cal13,mal22}. Also, recent 1D simulations for main sequence stars by \citet{oet20}, suggest that the dynamical interaction between an MS star (in the range $0.8-1.0$~\msun) and its companion ($1-20$~M$_J$) can induce an increase of the rotational velocity of the host star. This occurs while it is still on the main sequence from 1 to over 40~km~s$^{-1}$ if an engulfment occurs, depending on the initial semi-major axis, the mass ratio planet to star, and the metallicity of the host star.

We did not include any initial intrinsic rotation of the colliding bodies in our simulations. A rotation such as that of the present Sun ($\sim 2$~km~s$^{-1}$) is extremely small in comparison with the highly dynamic interaction of the stars. For the longest simulations (mergers), which are on the order of several $10^5$~s, the outer layers of the MS star would have rotated less than $10^6$~km. This is three orders of magnitude smaller than the spatial resolution in the lower-density regions of the MS star. The rotational profile of a BD is less known and high rotational velocities may be expected ($v \sin i>10$~km~s$^{-1}$, see e.g. \cite{konopacky2012}). However, these values are still small to produce significant changes within dynamical timescales. Nevertheless, we note that three extremely fast rotating BDs have been recently observed \citep{tannock2021}; one of them with the highest $v \sin i\sim 100$~km~s$^{-1}$ value ever reported for an ultra-cool BD.

\begin{figure}
\includegraphics[width=\columnwidth, center]{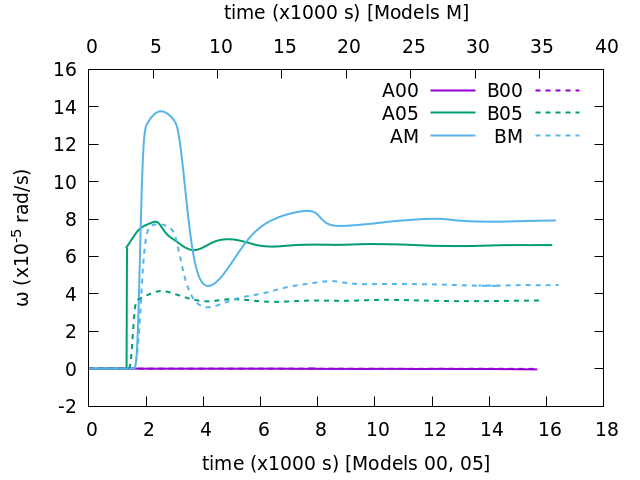}
    \caption{Average angular velocity evolution for the A (solid lines) and B (dashed lines) models. The calculation is restricted to the fluid within $R=1$~\rsun.}
    \label{fig:angvel}
\end{figure}

We measured the evolution of the angular momentum of the fluid within 1~\rsun~ from the corresponding center of mass for all simulated models. From that, and calculating the moment of inertia, we can derive an average angular velocity of the fluid at all times. We restrict the calculation to a sphere with radius $r=1$~\rsun, which will provide a lower boundary to the final angular velocity, as we expect this to increase when the expelled (but still gravitationally bounded) matter will fall back onto the host star. We also conducted the same analysis with a larger radius ($r=1.2$~\rsun) which yielded similar results. Figure~\ref{fig:angvel} shows the evolution of the angular velocity for all simulations and Table~\ref{table_results} shows the corresponding final period of the MS star. As expected, the head-on cases produce almost no change in the rotation, while grazing collisions and mergers have a remarkable effect. The first peak and subsequent dip in the evolution of the mergers for $t=5000-12500$~s is a direct consequence of how we calculate the angular momentum. Because we restrict it to $r=1$~\rsun, the value changes drastically when the BD crosses that limit, in and out, during its first orbit. After that, the trajectory of the accreted BD material remains always within that radius and the evolution of the angular velocity is smoother (increasing slowly as material keeps being accreted). Model A05 increases the rotational velocity of the MS star up to 46~km~s$^{-1}$, while model AM up to $\sim 52$~km~s$^{-1}$ at the end of the simulation.
For a given choice of the initial relative velocity and impact parameter (Table~\ref{table_models}), this increment factor would be similar for a similar mass ratio between the BD and the MS star, in agreement with the 1D simulations by \citet{oet20} (see Sect.~\ref{sec:changing_mass}). Regarding the radial profile of the angular velocity for the final objects of models A05 and AM, we found that they are in a transitory situation. The core of the MS star ($\sim 0.2$~\msun) exhibits a strong differential rotation, while the remainder of the star shows a rotation closer to rigid body, increasing only in the outer layers, where mass is still accreting. The fact that the average angular velocity shown in Fig.~\ref{fig:angvel} is stable for most of the last part of all simulations despite its radial variability, suggests a constant redistribution of angular momentum so that the provided angular velocity is a possible observable. Additionally, the fast rotating central core might have observational implications for asteroseismology studies through the detection of $g$-modes in those stars suspected to have recently suffered a sub-stellar engulfment \citep[see e.g.][]{fos17,egg19}.
\begin{figure*}
\centering
\includegraphics[width=\textwidth]{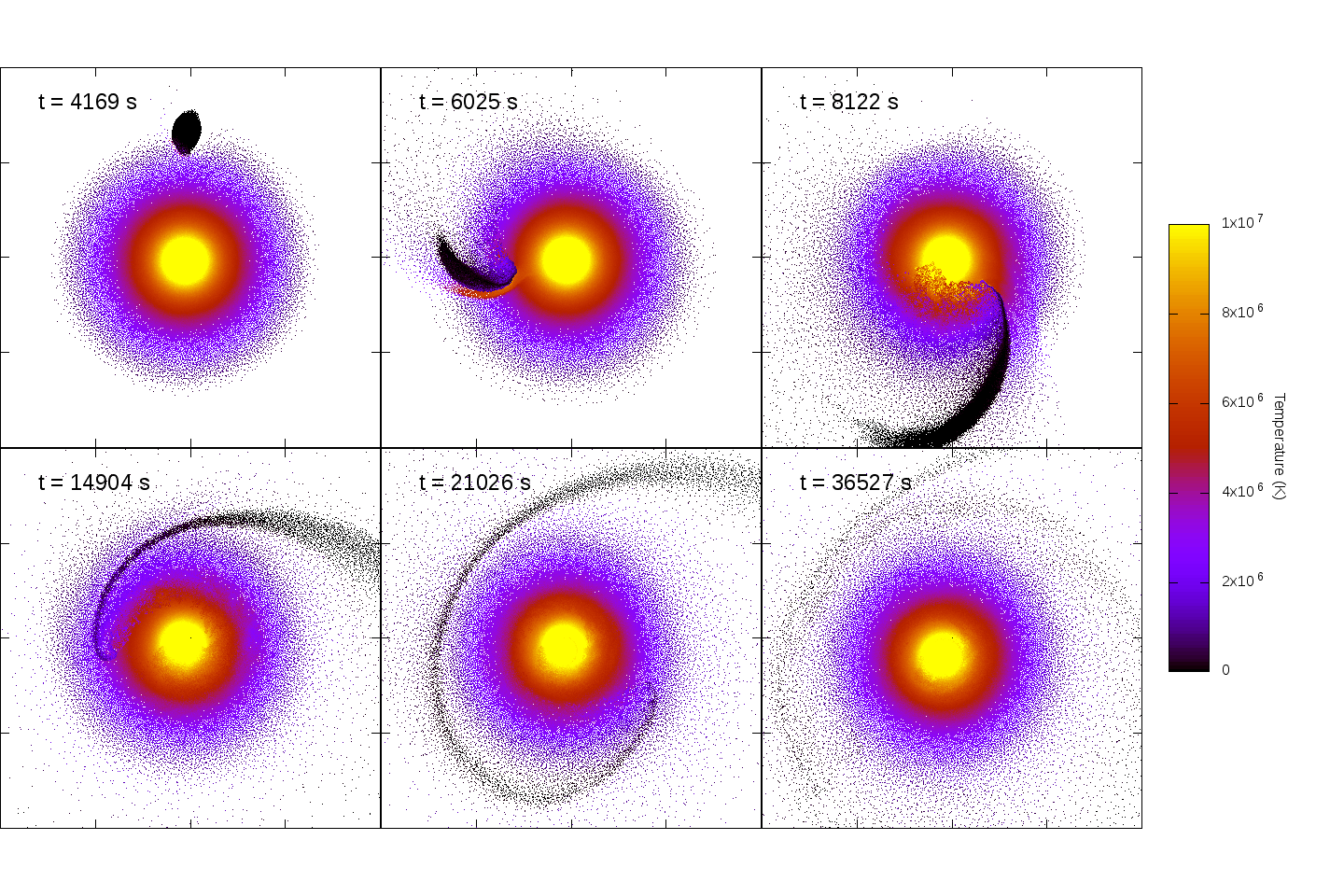}
    \caption{Snapshots of the particle distribution for the merger scenario with the lower mass BD (model BM) at different times. We only show particles within a thin slice along the orbital plane. Each snapshot shows a box with side $2\times10^{11}$~cm and temperature is color-coded.}
    \label{fig:snapshots_bm}
\end{figure*}

Current observational measurements of the rotational velocity in field MS star samples are still small to fully test our model. Nevertheless, the values obtained can be compared with the equatorial angular velocity of the Sun ($\sim 2$~km~s$^{-1}$) and with the $v\sin i$ values measured in a few main sequence stars of similar spectral type suspected to have suffered a planet engulfment and/or the accretion of rocky material. For instance, it has been found that a significant fraction among a sample of F stars hosting HJ show average $v\sin i$ values larger than those stars in their sample (in the same effective temperature region) without HJ \citep{del15}. These moderate rotators show in addition, lower Li abundances (by 0.14~dex), suggesting that rotational mixing (induced by planet interaction/engulfment?) might be the cause for a greater depletion of Li. On the contrary, in the 16 Cyg pair, star A (with a Li excess supposedly due to the accretion of some rocky material) shows a slightly larger rotational velocity (2.23~km~s$^{-1}$) than star B (1.27~km~s$^{-1}$) \citep{dav15}. Similar values are observed in the co-moving pair Kronos and Krios: the star HD 240430, with Li abundance $\sim +0.5$~dex larger, shows also a larger rotational velocity that HD 240429 (2.5 vs. 1.1~km~s$^{-1}$, \citet{oh18}).

All these observed rotational velocity values are compatible with that expected for main sequence stars of their masses and age \citep{sku72,tass00}, and contrast with the more than one order of magnitude increase in the angular velocity expected from interactions such as those presented here, in models A. The  observations existing so far do not show main-sequence field stars with mass and age similar to the Sun, with $v\sin i\gtrsim 10$~km~s$^{-1}$ \citep{don96,sta04}. Such large rotational velocities are only observed in some solar-like main sequence stars belonging to very young open clusters (e.g. in the Pleiades, \citet{que98,ter00,reb16}; Praesepe, \citet{dou19}; or NGC 6811, \citet{cur19}). Note however, that for sun-like cool stars the torque triggered by the magnetic braking mechanism increases with the increasing angular velocity implying, as a consequence, a spin-down rate $\dot{\Omega}\sim (\Omega_o/2t_o)(\Omega_\star/\Omega_o)^3$; with $t_o= 2\times 10^8$ a and $\Omega_o=10$~km~s$^{-1}/R_\star$ \citep[see e.g.][]{sku72,sod83,pan18}. Therefore, for R$_\star\sim 1$ R$_\sun$ and the surface $\Omega_\star$ values quoted above, the rotational velocity would be reduced in a time considerably lower than $\sim 10^9$ a. This would make difficult the observational detection of any rotational velocity excess induced by an engulfment like those studied here, considering the relatively low frequency of these interactions. Nevertheless, there is observational evidence for a possible dependence on the stellar mass of the braking law, resulting in K dwarfs spinning down much more slowly than F and G dwarfs, and that this braking may stall for an extended period of time, the duration of which increases toward lower masses/cooler temperatures \citep[e.g.][]{cur19}.
In any case, one should wait for more extensive and dedicated surveys on the rotational velocity in solar-like field stars before a definite conclusion is reached.

\subsection{Changing the mass of the BD}
\label{sec:changing_mass}

\begin{figure*}
\includegraphics[width=\textwidth]{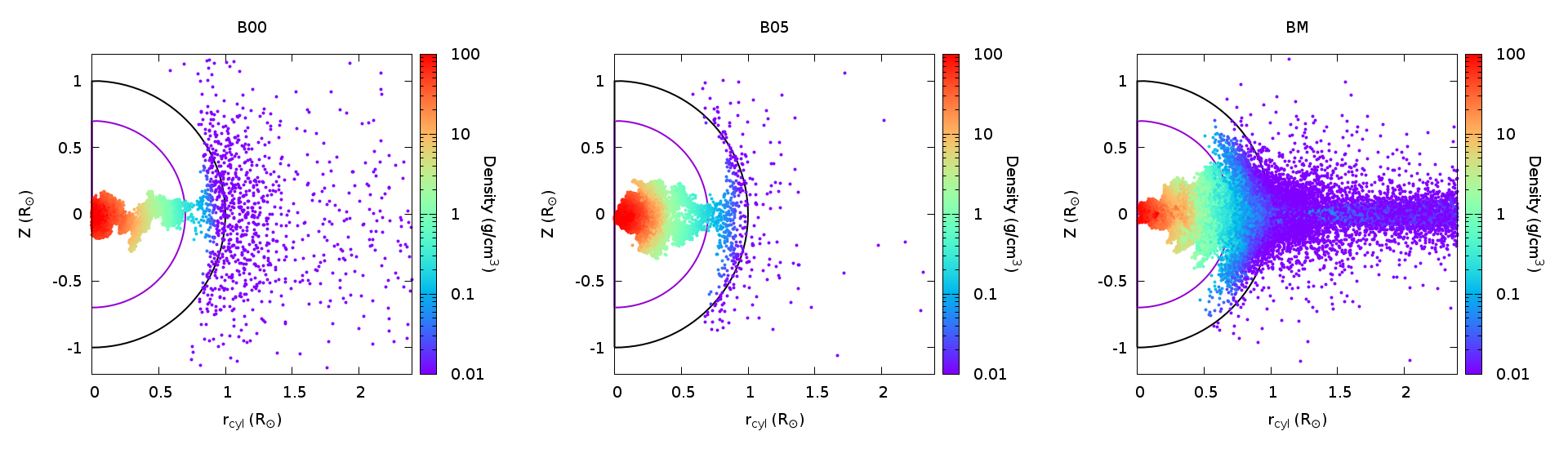}
    \caption{Particle distribution of the BD projected in the 2D cylindrical plane for the scenarios B, at the end of the simulations: $t=11478$~s and $t=11923$~s in the collision cases (left and central, respectively) and $t=29821$~s for the merger scenario (right). The semicircles represent the original solar radius (solid black line) and the bottom of the convective zone at $\sim 70\%$ of the solar radius (purple solid line). Density is color-coded.}
    \label{fig:small_BD_dist}
\end{figure*}
To explore the impact of the mass of the engulfed star, we reduced the mass of the BD adopted in models A ($0.019$~\msun) to $0.01$~\msun~in models B (i.e. closer to a giant planet mass), and we re-calculated the same three interactions with the same MS stellar mass and initial configurations. The relaxation process for this lighter BD was the same as for models A, and it uses the same EOS assuming the same Li abundance. In Fig.~\ref{fig:initialmodels}, right panel, we show the initial density profile of the 3D particle distribution after relaxation, compared with the original 1D theoretical profile. Being a factor 2.8 less dense, this lighter BD is expected to be destroyed more easily than the heavier one.

In the case of the head-on and grazing models (B00 and B05), the dynamics are similar to those already discussed regarding models A00 and A05. The main difference is that the BD is destroyed faster by its interaction with the MS star, as expected. Collision models B also eject $3-4$ times less mass than collision models A (see Table~\ref{table_results}) owing to the lower kinetic energy stored in the less massive BD at the time of the impact. Interestingly, model B00 is able to eject 3 times more mass from the BD than model A00, most likely due to its lower compactness, which makes the ablation process more efficient. The case of the merger (BM) also presents a similar evolution to that of model AM (see Fig.~\ref{fig:snapshots_bm}). The BD is again torn apart by tidal forces in half of an orbit, but the resulting spiral arm is more diluted in this case and it mixes rapidly with the surrounding material of the MS star. Overall, the merger produces a similar amount of ejected mass as in case A, belonging exclusively to the MS in both cases.

The BD mass distributions at the end of the simulations (see Fig.~\ref{fig:small_BD_dist}) follow the same pattern as in models A. In fact, the fraction of the BD mass which remains in the outer layers of the MS star is very similar to that in models A (see Table~\ref{table_results}). This can also be seen in the final periods of the MS star, which follow the same hierarchy as in models A, albeit with slightly lower angular velocities due to the lower total angular momentum of the system with the smaller BD.

Figure~\ref{fig:BDmass_07rsun} shows the time evolution of the percentage of the BD mass with $r\geq 0.7$~\rsun~for cases B with dashed lines. In these cases, the results show that the material of the BD can be ablated more easily when the BD mass is lower, as expected because of its lower gravitational energy content. Figure~\ref{fig:small_BD_dist} shows the cylindrical projection of all BD particles at the end of each simulation. Comparing it with Fig.~\ref{fig:BD_dist}, the same conclusions as for models A  apply here. The main difference appears in model BM with respect to model AM, as the BD mass in the outer regions is distributed in a thinner disk than in case AM, more concentrated around the collision plane but still with similar density in the equator. Therefore, similarly to model AM, in model BM also a circumstellar disk of matter around the MS star might form that eventually may produce an infrared excess. We suggest performing dedicated studies for the detection of infrared excess in MS stars with anomalous Li abundances. We note that large infrared excesses have been found in K-giants with enhanced Li abundances, which in addition are usually fast rotators \citep{reb15}, although the actual origin of the Li enhancement is still unclear \citep{zho19}.

As in models A, it is straightforward to estimate the new Li abundance in the envelope of the MS star at the end of the model B simulations. We obtain A(Li)$\sim 1.31, 1.13$, and $2.24$ for models B00, B05, and BM, respectively. These values are very similar to those obtained in models A. Therefore, the same conclusions on the observational detectability of this Li increase mentioned in Sect. 3.2 hold in the case of a lower BD mass. Although the fraction of the BD mass deposited in the envelope is larger in models B, the net mass deposited is lower.

Regarding the angular momentum transfer to the MS star, Fig.~\ref{fig:angvel} shows the time evolution of the angular velocity of the fluid with $r\leq$ \rsun. It is not surprising that there is a dependency on the mass of BD. The factor between the final angular velocity (see Table~\ref{table_results}) of models with different mass is $\sim 2$, which is very similar to the mass ratio of the BDs of models A and B. Finally, the same trend observed in models A between types of collisions and final angular velocity is observed in models B. Owing that very little BD mass is lost by the system (see Table~\ref{table_results}, third column) a linear relationship between final angular velocity and mass of the impacting body is expected. Therefore, an impact similar to those simulated here, with a Jupiter-mass planet instead of a BD, should lead to angular velocities about 8 times lower than cases B just after the event takes place. Namely, $v\sin i\sim 3.1$~km~s$^{-1}$ for a grazing collision with an impact parameter of $0.5$~\rsun, and $v\sin i\sim 3.6 $~km~s$^{-1}$ for a merger-like scenario. These rotational velocities are of the same order as those typically observed in solar-like stars. Therefore, the rotational velocity would not serve as an observational hint to detect singular collisions/engulfments of a planetary mass similar to that of Jupiter or less by a solar-like MS star.

Finally, the fate of the Li-rich material has a very similar trend as in models A. As Fig.~\ref{fig:ttmaxTemp} (second row) shows, all BD material with $r\lesssim0.7$~\rsun~will be inexorably burnt. Only the material on the outer layers never reaches the Li-burning temperature or, if it does, it is for a short time span. 

Looking at Fig.~\ref{fig:ttmaxTemp} it is interesting to note that a clear boundary appears naturally in all simulations between burnt and unburnt material at $r\sim0.7$~\rsun, which is traditionally assumed as the location of the bottom of the convective envelope. No previous assumption was made in this respect in our code, yet the boundary appeared naturally, independently of the mass of BD. Despite the highly dynamic interaction and that the final body is still not in equilibrium at the end of the simulations, this result reinforces our premise that the location of the bottom of the convective envelope will not change considerably.

\section{Summary and conclusions}
\label{sec:conclusions}
In this work, we have explored the effects of the engulfment of a brown dwarf by a solar-like main-sequence star. Our main goal is to study if this interaction may significantly alter the lithium abundance on the surface of the host star and induce changes in its main properties with possible observational consequences. To do so, we conducted  3D hydrodynamical simulations exploring different types of interaction (head-on and grazing collisions, and mergers) and the impact of changing the BD mass. Our main findings are summarized in the following:
\begin{itemize}

    \item In all the explored scenarios, the majority of the BD mass remains in the inner parts of the MS star, even reaching the stellar core. This is at odds with previous suggestions based on 1D simulations, that BD with masses $\leq 0.02$~\msun~would mostly dilute in the envelope of the MS star. In our $M_{BD}=0.019$~\msun~case, collisions produce very little pollution in the outer shells of the MS star ($\sim 0.3-1.4\%$ of the BD mass). Even with zero impact parameter, where a low-pressure funnel eases the flow of BD material to external layers, almost all BD material will be confined in the high-density region of the host star. When decreasing the mass of the BD to $0.01$~\msun, the percent of its mass that remains in the outer layers increases only slightly ($\sim 0.9-2.4\%$). Thus, on the basis of the present simulations, we can not elucidate if a complete dilution of the impacting object in the MS star's envelope would occur at much lower masses, such as that of a giant planet.

    \item In the case of mergers, a sizeable fraction ($\simeq 40\%$ for both models) of the BD mass settles in the outer layers of the MS star and never reaches Li-burning temperatures.

    \item We find a Li enhancement in the envelope of the MS star at the end of the simulated interactions. Such enhancement is very small in the head-on and grazing collision scenarios for the two BD masses studied. Nevertheless, the enhancement would be close to a factor of $20-30$ in the case of a merger. In this latter case, assuming that the enrichment remains long enough in the envelope, it could be observationally detected as a Li enhancement in surveys of solar-like MS stars. It is worth noting that these values are indicative, as the final state in our simulations is still evolving (albeit in a timescale longer than the dynamical scale). Therefore, additional work should be done exploring the long-term evolution of the system.

    \item There is a clear increase in the rotational velocity of the MS star after the interaction, being larger in the case of a merger ($\sim 52$~km~s$^{-1}$ for AM, and $\sim 29$~km~s$^{-1}$ for BM), although quite similar to that produced by a grazing collision ($\sim 46$~km~s$^{-1}$ for A05, and $\sim 25$~km~s$^{-1}$ for B05). This signal is nevertheless transitory. Even if the rotational velocity can increase even more due to the fallback of the MS and BD material that is still gravitationally bound, mechanisms such as magnetic breaking would spin-down the star in short timescales ($\lesssim 10^9$~a).

    \item As most of the BD mass remains gravitationally bound to the MS star after the interaction, the angular momentum transfer is directly proportional to the mass of the BD and the dynamical initial conditions. Assuming a similar kinetic content and type of impact, this allows us to extrapolate the rotational velocity of the host star in the case of an interaction with a Jupiter-like mass planet, being in the range of $2-5$~km~s$^{-1}$, depending on the type of interaction. This would discard rotational velocity as a proxy to select solar-like MS stars that may have suffered an engulfment of a planet-mass object. High rotational velocity values might be used instead as a probe to identify planet/BD interactions with red giant stars as shown by \citet{pri16,pri16b}. This will be the subject of a similar forthcoming study.

    \item A significant amount of mass is lost from the system (in the range $10^{-4}-10^{-3}$~\msun) in all our simulations. The overwhelming majority of this mass belongs to the host star. In the case of a merger, this mass may form a disk of circumstellar matter, which may have observational consequences such as an infrared excess. This agrees with previous results from 1D studies on the interaction of a substellar object with a MS  \citep[e.g.,][]{ste20} and/or giant star \citep[e.g.,][]{sok98,sok01}.



\end{itemize}

It has to be emphasized that, although the evolving timescale at the end of the simulation is much larger than the dynamical timescale, the final fluid distribution is still not in complete equilibrium. To follow the simulation until all gravitationally-bound matter falls back and the final object reaches a stable configuration is out of reach for a full 3D hydrodynamical simulation limited by the Courant condition. To know the final distribution of matter, a stellar evolution code should be used. Nevertheless, we want to stress the asymmetric nature of the simulated interactions and their lack of hydrostatic equilibrium. In order to follow its evolution, we would need a 3D hydrodynamical evolution code, which nowadays is not available.

In summary, our 3D simulations show that the outcome of the engulfment of a sub-stellar object by an MS star depends on the geometry of the engulfment and the mass of the colliding object. The main differences are the induced rotational velocity and the amount of mass of the sub-stellar object which is whether diluted in the outer layers or in the interior of the MS star. As a result, some of these interactions yield a considerable mass loss and a measurable change in the surface Li abundance, both of which could be observationally detected.

\section*{Acknowledgements}
This work has been supported by the European Research Council (FP7) under ERC Advanced Grant Agreement No. 321263 - FISH, by the Swiss Platform for Advanced Scientific Computing (PASC) projects DIAPHANE and SPH-EXA, and by the SKACH consortium through funding by SERI. It has also been supported by the Spanish projects PGC2018-095317-B-C21,  PID2021-123110NB-100 financed by the MCIN/AEI FEDER “Una manera de hacer Europa”, and by the MINECO Spanish project PID2020-117252GB-100. The authors acknowledge and thank the invaluable support of the scientific computing core facility sciCORE at the University of Basel (\url{https://scicore.unibas.ch/}), the Swiss National Supercomputing Center CSCS (\url{https://cscs.ch}), the Centro de Servicios Informáticos y Redes de Comunicación CSiRC at the University of Granada (\url{https://csirc.ugr.es/}), and the Instituto de Astrofísica de Canarias (\url{https://www.iac.es/}). All these institutions provided infrastructures and resources to perform these calculations.

\bibliographystyle{aa}
\bibliography{bibliography}

\begin{thebibliography}{103}
\expandafter\ifx\csname natexlab\endcsname\relax\def\natexlab#1{#1}\fi

\bibitem[{{Agertz} {et~al.}(2007){Agertz}, {Moore}, {Stadel}, {Potter},
  {Miniati}, {Read}, {Mayer}, {Gawryszczak}, {Kravtsov}, {Nordlund}, {Pearce},
  {Quilis}, {Rudd}, {Springel}, {Stone}, {Tasker}, {Teyssier}, {Wadsley}, \&
  {Walder}}]{agertz2007}
{Agertz}, O., {Moore}, B., {Stadel}, J., {et~al.} 2007, \mnras, 380, 963

\bibitem[{{Aguilera-G{\'o}mez} {et~al.}(2020){Aguilera-G{\'o}mez},
  {Chanam{\'e}}, \& {Pinsonneault}}]{agu20}
{Aguilera-G{\'o}mez}, C., {Chanam{\'e}}, J., \& {Pinsonneault}, M.~H. 2020,
  \apjl, 897, L20

\bibitem[{{Aguilera-G{\'o}mez} {et~al.}(2016){Aguilera-G{\'o}mez},
  {Chanam{\'e}}, {Pinsonneault}, \& {Carlberg}}]{agu16}
{Aguilera-G{\'o}mez}, C., {Chanam{\'e}}, J., {Pinsonneault}, M.~H., \&
  {Carlberg}, J.~K. 2016, \apj, 829, 127

\bibitem[{{Alexander}(1967)}]{ale67}
{Alexander}, J.~B. 1967, The Observatory, 87, 238

\bibitem[{{Asplund} {et~al.}(2009){Asplund}, {Grevesse}, {Sauval}, \&
  {Scott}}]{aps09}
{Asplund}, M., {Grevesse}, N., {Sauval}, A.~J., \& {Scott}, P. 2009, \araa, 47,
  481

\bibitem[{{Bahcall}(2005)}]{bah05}
{Bahcall}, J.~N. 2005, Physica Scripta Volume T, 121, 46

\bibitem[{{Bahcall} {et~al.}(2001){Bahcall}, {Pinsonneault}, \& {Basu}}]{bah01}
{Bahcall}, J.~N., {Pinsonneault}, M.~H., \& {Basu}, S. 2001, \apj, 555, 990

\bibitem[{{Baraffe} \& {Chabrier}(2010)}]{bar10}
{Baraffe}, I. \& {Chabrier}, G. 2010, \aap, 521, A44

\bibitem[{{Baumann} {et~al.}(2010){Baumann}, {Ram{\'\i}rez}, {Mel{\'e}ndez},
  {Asplund}, \& {Lind}}]{bau10}
{Baumann}, P., {Ram{\'\i}rez}, I., {Mel{\'e}ndez}, J., {Asplund}, M., \&
  {Lind}, K. 2010, \aap, 519, A87

\bibitem[{{Bear} \& {Soker}(2011)}]{bea11}
{Bear}, E. \& {Soker}, N. 2011, \mnras, 414, 1788

\bibitem[{{Burrows} \& {Liebert}(1993)}]{burrows1993}
{Burrows}, A. \& {Liebert}, J. 1993, Reviews of Modern Physics, 65, 301

\bibitem[{{Cabez{\'o}n} {et~al.}(2012){Cabez{\'o}n}, {Garc{\'{\i}}a-Senz}, \&
  {Escart{\'{\i}}n}}]{cabezon2012}
{Cabez{\'o}n}, R.~M., {Garc{\'{\i}}a-Senz}, D., \& {Escart{\'{\i}}n}, J.~A.
  2012, \aap, 545, A112

\bibitem[{{Cabez{\'o}n} {et~al.}(2017){Cabez{\'o}n}, {Garc{\'{\i}}a-Senz}, \&
  {Figueira}}]{cabezon2017}
{Cabez{\'o}n}, R.~M., {Garc{\'{\i}}a-Senz}, D., \& {Figueira}, J. 2017, \aap,
  606, A78

\bibitem[{{Cabez{\'o}n} {et~al.}(2008){Cabez{\'o}n}, {Garc{\'{\i}}a-Senz}, \&
  {Rela{\~n}o}}]{cabezon2008}
{Cabez{\'o}n}, R.~M., {Garc{\'{\i}}a-Senz}, D., \& {Rela{\~n}o}, A. 2008,
  Journal of Computational Physics, 227, 8523

\bibitem[{{Carlberg} {et~al.}(2013){Carlberg}, {Cunha}, {Smith}, \&
  {Majewski}}]{cal13}
{Carlberg}, J.~K., {Cunha}, K., {Smith}, V.~V., \& {Majewski}, S.~R. 2013,
  Astronomische Nachrichten, 334, 120

\bibitem[{{Carlberg} {et~al.}(2009){Carlberg}, {Majewski}, \& {Arras}}]{car09}
{Carlberg}, J.~K., {Majewski}, S.~R., \& {Arras}, P. 2009, \apj, 700, 832

\bibitem[{{Charbonnel} \& {Talon}(2005)}]{char05}
{Charbonnel}, C. \& {Talon}, S. 2005, Science, 309, 2189

\bibitem[{{Chen} \& {Zhao}(2006)}]{che06}
{Chen}, Y.~Q. \& {Zhao}, G. 2006, \aj, 131, 1816

\bibitem[{{Cullen} \& {Dehnen}(2010)}]{cullen2010}
{Cullen}, L. \& {Dehnen}, W. 2010, \mnras, 408, 669

\bibitem[{{Curtis} {et~al.}(2019){Curtis}, {Ag{\"u}eros}, {Douglas}, \&
  {Meibom}}]{cur19}
{Curtis}, J.~L., {Ag{\"u}eros}, M.~A., {Douglas}, S.~T., \& {Meibom}, S. 2019,
  \apj, 879, 49

\bibitem[{{D'Antona} \& {Mazzitelli}(1994)}]{dan94}
{D'Antona}, F. \& {Mazzitelli}, I. 1994, \apjs, 90, 467

\bibitem[{{D'Antona} \& {Montalb{\'a}n}(2003)}]{dan03}
{D'Antona}, F. \& {Montalb{\'a}n}, J. 2003, \aap, 412, 213

\bibitem[{{Davies} {et~al.}(2015){Davies}, {Chaplin}, {Farr}, {Garc{\'\i}a},
  {Lund}, {Mathis}, {Metcalfe}, {Appourchaux}, {Basu}, {Benomar}, {Campante},
  {Ceillier}, {Elsworth}, {Handberg}, {Salabert}, \& {Stello}}]{dav15}
{Davies}, G.~R., {Chaplin}, W.~J., {Farr}, W.~M., {et~al.} 2015, \mnras, 446,
  2959

\bibitem[{{Deal} {et~al.}(2015){Deal}, {Richard}, \& {Vauclair}}]{ddd15}
{Deal}, M., {Richard}, O., \& {Vauclair}, S. 2015, \aap, 584, A105

\bibitem[{{Delgado Mena} {et~al.}(2015){Delgado Mena}, {Bertr{\'a}n de Lis},
  {Adibekyan}, {Sousa}, {Figueira}, {Mortier}, {Gonz{\'a}lez Hern{\'a}ndez},
  {Tsantaki}, {Israelian}, \& {Santos}}]{del15}
{Delgado Mena}, E., {Bertr{\'a}n de Lis}, S., {Adibekyan}, V.~Z., {et~al.}
  2015, \aap, 576, A69

\bibitem[{{Donahue} {et~al.}(1996){Donahue}, {Saar}, \& {Baliunas}}]{don96}
{Donahue}, R.~A., {Saar}, S.~H., \& {Baliunas}, S.~L. 1996, \apj, 466, 384

\bibitem[{{Douglas} {et~al.}(2019){Douglas}, {Curtis}, {Ag{\"u}eros},
  {Cargile}, {Brewer}, {Meibom}, \& {Jansen}}]{dou19}
{Douglas}, S.~T., {Curtis}, J.~L., {Ag{\"u}eros}, M.~A., {et~al.} 2019, \apj,
  879, 100

\bibitem[{{Eggenberger} {et~al.}(2019){Eggenberger}, {Buldgen}, \&
  {Salmon}}]{egg19}
{Eggenberger}, P., {Buldgen}, G., \& {Salmon}, S.~J.~A.~J. 2019, \aap, 626, L1

\bibitem[{{Fossat} {et~al.}(2017){Fossat}, {Boumier}, {Corbard}, {Provost},
  {Salabert}, {Schmider}, {Gabriel}, {Grec}, {Renaud}, {Robillot},
  {Roca-Cort{\'e}s}, {Turck-Chi{\`e}ze}, {Ulrich}, \& {Lazrek}}]{fos17}
{Fossat}, E., {Boumier}, P., {Corbard}, T., {et~al.} 2017, \aap, 604, A40

\bibitem[{{Frontiere} {et~al.}(2017){Frontiere}, {Raskin}, \&
  {Owen}}]{frontiere2017}
{Frontiere}, N., {Raskin}, C.~D., \& {Owen}, J.~M. 2017, Journal of
  Computational Physics, 332, 160

\bibitem[{{Galarza} {et~al.}(2021){Galarza}, {L{\'o}pez-Valdivia},
  {Mel{\'e}ndez}, \& {Lorenzo-Oliveira}}]{yan21}
{Galarza}, J.~Y., {L{\'o}pez-Valdivia}, R., {Mel{\'e}ndez}, J., \&
  {Lorenzo-Oliveira}, D. 2021, \apj, 922, 129

\bibitem[{{Garc{\'{\i}}a-Senz} {et~al.}(2012){Garc{\'{\i}}a-Senz},
  {Cabez{\'o}n}, \& {Escart{\'{\i}}n}}]{garcia2012}
{Garc{\'{\i}}a-Senz}, D., {Cabez{\'o}n}, R.~M., \& {Escart{\'{\i}}n}, J.~A.
  2012, \aap, 538, A9

\bibitem[{{Garc{\'\i}a-Senz} {et~al.}(2022){Garc{\'\i}a-Senz}, {Cabez{\'o}n},
  \& {Escart{\'\i}n}}]{gar2022}
{Garc{\'\i}a-Senz}, D., {Cabez{\'o}n}, R.~M., \& {Escart{\'\i}n}, J.~A. 2022,
  \aap, 659, A175

\bibitem[{{Gonzalez}(2000)}]{gon00}
{Gonzalez}, G. 2000, in Astronomical Society of the Pacific Conference Series,
  Vol. 219, Disks, Planetesimals, and Planets, ed. G.~{Garz{\'o}n}, C.~{Eiroa},
  D.~{de Winter}, \& T.~J. {Mahoney}, 523

\bibitem[{{Gonzalez}(2015)}]{gon15}
{Gonzalez}, G. 2015, \mnras, 446, 1020

\bibitem[{{Gonzalez} {et~al.}(2010){Gonzalez}, {Carlson}, \& {Tobin}}]{gon10}
{Gonzalez}, G., {Carlson}, M.~K., \& {Tobin}, R.~W. 2010, \mnras, 407, 314

\bibitem[{{Greenstein} \& {Richardson}(1951)}]{gre51}
{Greenstein}, J.~L. \& {Richardson}, R.~S. 1951, \apj, 113, 536

\bibitem[{{Hopkins}(2013)}]{hopkins2013}
{Hopkins}, P.~F. 2013, \mnras, 428, 2840

\bibitem[{{Israelian} {et~al.}(2009){Israelian}, {Delgado Mena}, {Santos},
  {Sousa}, {Mayor}, {Udry}, {Dom{\'\i}nguez Cerde{\~n}a}, {Rebolo}, \& {Rand
  ich}}]{isr09}
{Israelian}, G., {Delgado Mena}, E., {Santos}, N.~C., {et~al.} 2009, \nat, 462,
  189

\bibitem[{{Jackson} \& {Carlberg}(2018)}]{jac18}
{Jackson}, B. \& {Carlberg}, J. 2018, {Accretion of Planetary Material onto
  Host Stars} (Springer), 28

\bibitem[{{Jia} \& {Spruit}(2018)}]{jia18}
{Jia}, S. \& {Spruit}, H.~C. 2018, \apj, 864, 169

\bibitem[{{Kaiser} {et~al.}(2021){Kaiser}, {Clemens}, {Blouin}, {Dufour},
  {Hegedus}, {Reding}, \& {B{\'e}dard}}]{kai21}
{Kaiser}, B.~C., {Clemens}, J.~C., {Blouin}, S., {et~al.} 2021, Science, 371,
  168

\bibitem[{{King} {et~al.}(1997){King}, {Deliyannis}, {Hiltgen}, {Stephens},
  {Cunha}, \& {Boesgaard}}]{kin97}
{King}, J.~R., {Deliyannis}, C.~P., {Hiltgen}, D.~D., {et~al.} 1997, \aj, 113,
  1871

\bibitem[{{Konopacky} {et~al.}(2012){Konopacky}, {Ghez}, {Fabrycky},
  {Macintosh}, {White}, {Barman}, {Rice}, {Hallinan}, \&
  {Duch{\^e}ne}}]{konopacky2012}
{Konopacky}, Q.~M., {Ghez}, A.~M., {Fabrycky}, D.~C., {et~al.} 2012, \apj, 750,
  79

\bibitem[{{Kruckow} {et~al.}(2021){Kruckow}, {Neunteufel}, {Di Stefano}, {Gao},
  \& {Kobayashi}}]{kru21}
{Kruckow}, M.~U., {Neunteufel}, P.~G., {Di Stefano}, R., {Gao}, Y., \&
  {Kobayashi}, C. 2021, \apj, 920, 86

\bibitem[{{Lambert} \& {Reddy}(2004)}]{lam04}
{Lambert}, D.~L. \& {Reddy}, B.~E. 2004, \mnras, 349, 757

\bibitem[{{Liu} {et~al.}(2018){Liu}, {Yong}, {Asplund}, {Feltzing}, {Mustill},
  {Mel{\'e}ndez}, {Ram{\'\i}rez}, \& {Lin}}]{liu18}
{Liu}, F., {Yong}, D., {Asplund}, M., {et~al.} 2018, \aap, 614, A138

\bibitem[{{Livio} \& {Soker}(1984)}]{liv84}
{Livio}, M. \& {Soker}, N. 1984, \mnras, 208, 763

\bibitem[{{Lodders}(2019)}]{lod19}
{Lodders}, K. 2019, Solar Elemental Abundances, in The Oxford Research
  Encyclopedia of Planetary Science, Oxford University Press., arXiv:1912.00844

\bibitem[{{Luck} \& {Heiter}(2006)}]{luc00}
{Luck}, R.~E. \& {Heiter}, U. 2006, \aj, 131, 3069

\bibitem[{{MacLeod} {et~al.}(2018){MacLeod}, {Cantiello}, \&
  {Soares-Furtado}}]{mac18}
{MacLeod}, M., {Cantiello}, M., \& {Soares-Furtado}, M. 2018, \apjl, 853, L1

\bibitem[{{Maia} {et~al.}(2019){Maia}, {Mel{\'e}ndez}, {Lorenzo-Oliveira},
  {Spina}, \& {Jofr{\'e}}}]{mai19}
{Maia}, M.~T., {Mel{\'e}ndez}, J., {Lorenzo-Oliveira}, D., {Spina}, L., \&
  {Jofr{\'e}}, P. 2019, \aap, 628, A126

\bibitem[{{Mallick} {et~al.}(2022){Mallick}, {Reddy}, \&
  {Muthumariappan}}]{mal22}
{Mallick}, A., {Reddy}, B.~E., \& {Muthumariappan}, C. 2022, \mnras, 511, 3741

\bibitem[{{Metzger} {et~al.}(2012){Metzger}, {Rafikov}, \& {Bochkarev}}]{met12}
{Metzger}, B.~D., {Rafikov}, R.~R., \& {Bochkarev}, K.~V. 2012, \mnras, 423,
  505

\bibitem[{{Michaud}(1986)}]{mic86}
{Michaud}, G. 1986, \apj, 302, 650

\bibitem[{Mohammed {et~al.}(2020)Mohammed, Cavelan, Ciorba, Cabezón, \&
  Banicescu}]{mohammed2020}
Mohammed, A., Cavelan, A., Ciorba, F.~M., Cabezón, R.~M., \& Banicescu, I.
  2020, in Proceedings of the 2020 {SIAM} {Conference} on {Parallel}
  {Processing} for {Scientific} {Computing} ({PP}), Proceedings (Society for
  Industrial and Applied Mathematics), 69--80

\bibitem[{{Montalb{\'a}n} \& {D'Antona}(2006)}]{mon06}
{Montalb{\'a}n}, J. \& {D'Antona}, F. 2006, \mnras, 370, 1823

\bibitem[{{Montalb{\'a}n} \& {Rebolo}(2002)}]{mon02}
{Montalb{\'a}n}, J. \& {Rebolo}, R. 2002, \aap, 386, 1039

\bibitem[{{Nagar} {et~al.}(2020){Nagar}, {Spina}, \& {Karakas}}]{nag20}
{Nagar}, T., {Spina}, L., \& {Karakas}, A.~I. 2020, \apjl, 888, L9

\bibitem[{{Nordhaus} {et~al.}(2010){Nordhaus}, {Spiegel}, {Ibgui}, {Goodman},
  \& {Burrows}}]{nor10}
{Nordhaus}, J., {Spiegel}, D.~S., {Ibgui}, L., {Goodman}, J., \& {Burrows}, A.
  2010, \mnras, 408, 631

\bibitem[{{Oetjens} {et~al.}(2020){Oetjens}, {Carone}, {Bergemann}, \&
  {Serenelli}}]{oet20}
{Oetjens}, A., {Carone}, L., {Bergemann}, M., \& {Serenelli}, A. 2020, \aap,
  643, A34

\bibitem[{{Oh} {et~al.}(2018){Oh}, {Price-Whelan}, {Brewer}, {Hogg}, {Spergel},
  \& {Myles}}]{oh18}
{Oh}, S., {Price-Whelan}, A.~M., {Brewer}, J.~M., {et~al.} 2018, \apj, 854, 138

\bibitem[{{Pantolmos} \& {Matt}(2017)}]{pan18}
{Pantolmos}, G. \& {Matt}, S.~P. 2017, \apj, 849, 83

\bibitem[{{Piau} \& {Turck-Chi{\`e}ze}(2002)}]{pia02}
{Piau}, L. \& {Turck-Chi{\`e}ze}, S. 2002, \apj, 566, 419

\bibitem[{{Pinsonneault} {et~al.}(1992){Pinsonneault}, {Deliyannis}, \&
  {Demarque}}]{pin92}
{Pinsonneault}, M.~H., {Deliyannis}, C.~P., \& {Demarque}, P. 1992, \apjs, 78,
  179

\bibitem[{{Privitera} {et~al.}(2016{\natexlab{a}}){Privitera}, {Meynet},
  {Eggenberger}, {Vidotto}, {Villaver}, \& {Bianda}}]{pri16}
{Privitera}, G., {Meynet}, G., {Eggenberger}, P., {et~al.} 2016{\natexlab{a}},
  \aap, 591, A45

\bibitem[{{Privitera} {et~al.}(2016{\natexlab{b}}){Privitera}, {Meynet},
  {Eggenberger}, {Vidotto}, {Villaver}, \& {Bianda}}]{pri16b}
{Privitera}, G., {Meynet}, G., {Eggenberger}, P., {et~al.} 2016{\natexlab{b}},
  \aap, 593, A128

\bibitem[{{Queloz} {et~al.}(1998){Queloz}, {Allain}, {Mermilliod}, {Bouvier},
  \& {Mayor}}]{que98}
{Queloz}, D., {Allain}, S., {Mermilliod}, J.~C., {Bouvier}, J., \& {Mayor}, M.
  1998, \aap, 335, 183

\bibitem[{{Read} \& {Hayfield}(2012)}]{read2012}
{Read}, J.~I. \& {Hayfield}, T. 2012, \mnras, 422, 3037

\bibitem[{{Rebull} {et~al.}(2015){Rebull}, {Carlberg}, {Gibbs}, {Deeb},
  {Larsen}, {Black}, {Altepeter}, {Bucksbee}, {Cashen}, {Clarke}, {Datta},
  {Hodgson}, \& {Lince}}]{reb15}
{Rebull}, L.~M., {Carlberg}, J.~K., {Gibbs}, J.~C., {et~al.} 2015, \aj, 150,
  123

\bibitem[{{Rebull} {et~al.}(2016){Rebull}, {Stauffer}, {Bouvier}, {Cody},
  {Hillenbrand}, {Soderblom}, {Valenti}, {Barrado}, {Bouy}, {Ciardi},
  {Pinsonneault}, {Stassun}, {Micela}, {Aigrain}, {Vrba}, {Somers},
  {Christiansen}, {Gillen}, \& {Collier Cameron}}]{reb16}
{Rebull}, L.~M., {Stauffer}, J.~R., {Bouvier}, J., {et~al.} 2016, \aj, 152, 113

\bibitem[{{Rosswog}(2015)}]{rosswog2015}
{Rosswog}, S. 2015, \mnras, 448, 3628

\bibitem[{{Ryan}(2000)}]{rya00}
{Ryan}, S.~G. 2000, \mnras, 316, L35

\bibitem[{{Saitoh} \& {Makino}(2013)}]{saitoh2013}
{Saitoh}, T.~R. \& {Makino}, J. 2013, \apj, 768, 44

\bibitem[{{Sandquist} {et~al.}(1998){Sandquist}, {Taam}, {Lin}, \&
  {Burkert}}]{san98}
{Sandquist}, E., {Taam}, R.~E., {Lin}, D.~N.~C., \& {Burkert}, A. 1998, \apjl,
  506, L65

\bibitem[{{Sandquist} {et~al.}(2002){Sandquist}, {Dokter}, {Lin}, \&
  {Mardling}}]{san02}
{Sandquist}, E.~L., {Dokter}, J.~J., {Lin}, D.~N.~C., \& {Mardling}, R.~A.
  2002, \apj, 572, 1012

\bibitem[{{Siess} \& {Livio}(1999{\natexlab{a}})}]{sie99a}
{Siess}, L. \& {Livio}, M. 1999{\natexlab{a}}, \mnras, 304, 925

\bibitem[{{Siess} \& {Livio}(1999{\natexlab{b}})}]{sie99b}
{Siess}, L. \& {Livio}, M. 1999{\natexlab{b}}, \mnras, 308, 1133

\bibitem[{{Sills} {et~al.}(1997){Sills}, {Lombardi}, {Bailyn}, {Demarque},
  {Rasio}, \& {Shapiro}}]{Allison97}
{Sills}, A., {Lombardi}, James~C., J., {Bailyn}, C.~D., {et~al.} 1997, \apj,
  487, 290

\bibitem[{{Skumanich}(1972)}]{sku72}
{Skumanich}, A. 1972, \apj, 171, 565

\bibitem[{{Slattery} {et~al.}(1992){Slattery}, {Benz}, \&
  {Cameron}}]{slattery92}
{Slattery}, W.~L., {Benz}, W., \& {Cameron}, A.~G.~W. 1992, \icarus, 99, 167

\bibitem[{{Soares-Furtado} {et~al.}(2020){Soares-Furtado}, {Cantiello},
  {MacLeod}, \& {Ness}}]{soa20}
{Soares-Furtado}, M., {Cantiello}, M., {MacLeod}, M., \& {Ness}, M.~K. 2020,
  arXiv e-prints, arXiv:2002.05275

\bibitem[{{Soderblom}(1983)}]{sod83}
{Soderblom}, D.~R. 1983, \apjs, 53, 1

\bibitem[{{Soker}(1998)}]{sok98}
{Soker}, N. 1998, \aj, 116, 1308

\bibitem[{{Soker}(2001)}]{sok01}
{Soker}, N. 2001, \mnras, 324, 699

\bibitem[{{Spiegel} {et~al.}(2011){Spiegel}, {Burrows}, \& {Milsom}}]{spi11}
{Spiegel}, D.~S., {Burrows}, A., \& {Milsom}, J.~A. 2011, \apj, 727, 57

\bibitem[{{Staff} {et~al.}(2016){Staff}, {De Marco}, {Wood}, {Galaviz}, \&
  {Passy}}]{sta16}
{Staff}, J.~E., {De Marco}, O., {Wood}, P., {Galaviz}, P., \& {Passy}, J.-C.
  2016, \mnras, 458, 832

\bibitem[{{Stauffer}(2004)}]{sta04}
{Stauffer}, J.~R. 2004, in Stellar Rotation, ed. A.~{Maeder} \& P.~{Eenens},
  Vol. 215, 127

\bibitem[{{Stephan} {et~al.}(2020){Stephan}, {Naoz}, {Gaudi}, \&
  {Salas}}]{ste20}
{Stephan}, A.~P., {Naoz}, S., {Gaudi}, B.~S., \& {Salas}, J.~M. 2020, \apj,
  889, 45

\bibitem[{{Swenson} \& {Faulkner}(1992)}]{swe92}
{Swenson}, F.~J. \& {Faulkner}, J. 1992, \apj, 395, 654

\bibitem[{{Tannock} {et~al.}(2021){Tannock}, {Metchev}, {Heinze},
  {Miles-P{\'a}ez}, {Gagn{\'e}}, {Burgasser}, {Marley}, {Apai}, {Su{\'a}rez},
  \& {Plavchan}}]{tannock2021}
{Tannock}, M.~E., {Metchev}, S., {Heinze}, A., {et~al.} 2021, \aj, 161, 224

\bibitem[{{Tassoul}(2000)}]{tass00}
{Tassoul}, J.-L. 2000, {Stellar Rotation} (Cambridge University Press)

\bibitem[{{Terndrup} {et~al.}(2000){Terndrup}, {Stauffer}, {Pinsonneault},
  {Sills}, {Yuan}, {Jones}, {Fischer}, \& {Krishnamurthi}}]{ter00}
{Terndrup}, D.~M., {Stauffer}, J.~R., {Pinsonneault}, M.~H., {et~al.} 2000,
  \aj, 119, 1303

\bibitem[{{Th{\'e}ado} \& {Vauclair}(2012)}]{the12}
{Th{\'e}ado}, S. \& {Vauclair}, S. 2012, \apj, 744, 123

\bibitem[{{Timmes} \& {Swesty}(2000)}]{timmes2000}
{Timmes}, F.~X. \& {Swesty}, F.~D. 2000, \apjs, 126, 501

\bibitem[{{Umezu} \& {Saio}(2000)}]{ume00}
{Umezu}, M. \& {Saio}, H. 2000, \mnras, 316, 307

\bibitem[{{Valdarnini}(2016)}]{valdarnini2016}
{Valdarnini}, R. 2016, \apj, 831, 103

\bibitem[{{Villaver} \& {Livio}(2007)}]{vil07}
{Villaver}, E. \& {Livio}, M. 2007, \apj, 661, 1192

\bibitem[{{Villaver} {et~al.}(2014){Villaver}, {Livio}, {Mustill}, \&
  {Siess}}]{vil14}
{Villaver}, E., {Livio}, M., {Mustill}, A.~J., \& {Siess}, L. 2014, \apj, 794,
  3

\bibitem[{{Wang} {et~al.}(2021){Wang}, {Nordlander}, {Asplund}, {Amarsi},
  {Lind}, \& {Zhou}}]{wan21}
{Wang}, E.~X., {Nordlander}, T., {Asplund}, M., {et~al.} 2021, \mnras, 500,
  2159

\bibitem[{{Winn} \& {Fabrycky}(2015)}]{win15}
{Winn}, J.~N. \& {Fabrycky}, D.~C. 2015, \araa, 53, 409

\bibitem[{{Yamazaki} {et~al.}(2017){Yamazaki}, {Hayasaki}, \& {Loeb}}]{yam17}
{Yamazaki}, R., {Hayasaki}, K., \& {Loeb}, A. 2017, \mnras, 466, 1421

\bibitem[{{Zhou} {et~al.}(2019){Zhou}, {Yan}, {Shi}, {Blanco-Cuaresma}, {Gao},
  {Pan}, {Xu}, {Zhang}, \& {Zhao}}]{zho19}
{Zhou}, Y., {Yan}, H., {Shi}, J., {et~al.} 2019, \apj, 877, 104

\end{thebibliography}

\end{document}